\documentclass[a4paper,11pt]{article}
\pdfoutput=1 
\usepackage{jheppub} 
\usepackage{CJK}
\usepackage{natbib}

\usepackage[utf8]{inputenc}
\usepackage[american]{babel}
\usepackage[T1]{fontenc}
\usepackage{graphicx}
\usepackage{subfig}

\usepackage{pgf}
\usepackage{tikz}
\usepackage{amsmath}
\usepackage{nicefrac}
\usepackage{units}
\usepackage{todonotes}
\usepackage{mathcomp}
\usepackage{mathrsfs}
\usepackage{amssymb}
\usepackage{dsfont}
\usepackage{xcolor}
\usepackage{esvect}
\usepackage{slashed}
\usepackage{physics}
\usepackage{xspace}
\usepackage{bm}
\usepackage{wasysym}

\usepackage{cases}
\usepackage{hyperref}

\newcommand{\intd}{\textrm{d}}

\newcommand{\sigmatopmumu}{\Sigma^+\to p \mu^+\mu^-}
\newcommand{\sigmatopll}{\Sigma^+\to p \ell^+\ell^-}
\newcommand{\Bsigmatopmumu}{\mathcal{B}(\sigmatopmumu)} 
\newcommand{\Bsigmatopll}{\mathcal{B}(\sigmatopll)} 
\newcommand{\sigmatonlnu}{\Sigma^-\to n \ell^-\overline{\nu}_\ell}

\newcommand{\vect}[1]{\ensuremath{\bm{#1}}}

\newcommand{\AllLongRange}{Christ:2015pwa,Christ:2015aha,Christ:2016mmq,Briceno:2019opb,Boyle:2022ccj}
\newcommand{\gmtwopapers}{Bernecker:2011gh}

\begin{document}

\title{\boldmath Prospects for a lattice calculation of the rare decay \texorpdfstring{$\Sigma^+\rightarrow p\ell^+\ell^-$}{Sigma -> pl+l-}}
\author{Felix~Erben,}
\author{Vera~G\"ulpers,}
\author{Maxwell T.~Hansen,}
\author{Raoul~Hodgson,}
\author{Antonin~Portelli}
\affiliation{School of Physics and Astronomy, University of Edinburgh, Edinburgh EH9 3JZ, UK}
\date{\today}

\emailAdd{felix.erben@ed.ac.uk}
\emailAdd{vera.guelpers@ed.ac.uk}
\emailAdd{maxwell.hansen@ed.ac.uk}
\emailAdd{raoul.hodgson@ed.ac.uk}
\emailAdd{antonin.portelli@ed.ac.uk}

\abstract{We present a strategy for calculating the rare decay of a $\Sigma^+ (uus)$ baryon to a proton $(uud)$ and di-lepton pair using lattice QCD. To determine this observable one needs to numerically evaluate baryonic two-, three-, and four-point correlation functions related to the target process. In particular, the four-point function arises from the insertion of incoming and outgoing baryons, together with a weak Hamiltonian mediating the $s \to d$ transition and an electromagnetic current creating the outgoing leptons. As is described in previous work in other contexts, this four-point function has a highly non-trivial relation to the physical observable, due to proton and proton-pion intermediate states. These lead to growing Euclidean time dependence and, in the case of the two-particle intermediate states, to power-like volume effects. We discuss how to treat these issues in the context of the $\Sigma^+\rightarrow p\ell^+\ell^-$ decay and, in particular, detail the relation between the finite-volume estimator and the physical, complex-valued amplitude. In doing so, we also make connections between various approaches already described in the literature.}
\maketitle
\flushbottom



\section{Introduction}
\label{sec:intro}
The transition of an $s$- to a $d$-quark ($s\rightarrow d$) requires a \textit{flavour changing neutral current}, which is only allowed through quantum corrections within the Standard Model of
particle physics. Consequently, processes involving such transitions are rare in the Standard Model and could be
enhanced by potential new physics that includes a flavour changing neutral current in its Lagrangian. An example for such a
process is the rare semi-leptonic hyperon decay $\sigmatopll$, for which the muonic mode has been recently measured by the LHCb experiment
\cite{Aaij:2017ddf} with a branching ratio of
\begin{equation}
\Bsigmatopmumu = (2.2^{+1.8}_{-1.3})\times10^{-8}\,.
\label{eq:BRLHCb}
\end{equation}
Evidence for this decay had previously been found by the HyperCP collaboration \cite{Park:2005eka}
giving
\begin{equation}
\Bsigmatopmumu = (8.6^{^+6.6}_{-5.4}\pm5.5)\times 10^{-8} \,,
\label{eq:BRHyperCP}
\end{equation}
where the first uncertainties are statistical, and the second is systematic.
This determination follows from three events, all at nearly the same invariant mass of the $\mu^+\mu^-$ pair.
However, such a resonant structure in the $\mu^+\mu^-$ invariant mass could not be confirmed by the more
recent LHCb measurement \cite{Aaij:2017ddf}.

The current state-of-the-art Standard Model theory predictions for this process \cite{He:2005yn,He:2018yzu,Geng:2021fog} use a combination of dispersion relations, experimental input, various formulations of Baryon Chiral Perturbation Theory and model estimates (e.g.~vector meson dominance) and arrive at a range of results \cite{Geng:2021fog}
\begin{equation}
1.6\times 10^{-8}\leq \mathcal{B}\left(\sigmatopmumu\right) \leq 8.9\times10^{-8}\,.
\label{eq:BRHeetal}
\end{equation}
More details on the phenomenological background of this decay can be found in section \ref{sec:pheno}.

The rare hyperon decay $\sigmatopll$ can be viewed as the baryonic analogue of the rare kaon decay $K\to\pi\ell^+\ell^-$, which has been previously calculated from first principles using lattice simulations by the RBC/UKQCD collaboration, including recent results at physical quark masses \cite{Christ:2015aha,Christ:2016mmq,Boyle:2022ccj}. Taking inspiration from this progress, in this paper we explore prospects for calculating the required form factors for the $\sigmatopll$ decay using lattice QCD. While other hyperon decays with much higher yields, such as $\sigmatonlnu$, can be used to make measurements of the CKM matrix element $V_{us}$ in order to test for new physics that would break the CKM unitarity relations, this decay can be sensitive to new physics due to its rarity within the Standard Model.

As already discussed in refs.~\cite{\AllLongRange}, a key challenge in extracting decays such as $\sigmatopll$ and $K\to\pi\ell^+\ell^-$ from lattice QCD is that the physical observables (most directly defined in terms of infinite-volume Minkowski-signature correlation functions) contain on-shell intermediate states that can propagate between the weak Hamiltonian, effecting the $s \to d$ transition, and the electromagnetic current emitting the di-lepton pair. For the case of the rare hyperon decay in particular, intermediate $N \pi$ states contribute, where $N$ represents the nucleon doublet and $\pi$ the pion triplet. As we discuss in more detail in the following sections, in practice the states are projected to definite isospin as this is a good quantum number of the numerical calculation, provided the light quarks are degenerate and dynamical electromagnetism is not included, as we assume throughout. Fortunately, three-(or-more)-particle intermediate states are kinematically guaranteed to be off-shell and do not require special treatment.

Complications arise because numerical lattice QCD calculations only allow one to directly determine Euclidean correlation functions in a finite spacetime volume.
Specifically, the finite-volume Euclidean correlator that most closely matches the rare hyperon decay is a four-point function, defined with operators to create the incoming $\Sigma^+$ and the outgoing $p$ as well as the weak Hamiltonian and electromagnetic current.
Careful examination of this correlator shows that the on-shell $N \pi$ intermediate states manifest in a number of ways, all of which complicate the calculation.

First, after the baryonic operators are used to project out the $\Sigma^+$ and $p$ states, one finds that the on-shell intermediate states lead to exponentials that grow with the Euclidean-time separation between the weak Hamiltonian and the current \cite{Christ:2015pwa}. In practice, the number of such exponentials is finite, dictated by the discrete finite-volume spectrum, and thus these states can be removed through various strategies that we detail in the following sections.

However, a consequence of discarding these terms is that the resulting finite-volume estimator has poles at the locations of all finite-volume energies with $N\pi$ quantum numbers. In addition, the resulting quantity is known to have power-like volume effects away from the poles, and to miss the imaginary part appearing in the physical amplitude due to the long-distance propagation of intermediate states. In short, removing growing exponentials defines a finite-volume object that \emph{a priori} has no clear relation to the targeted amplitude.
Fortunately, as we describe in section \ref{sec:FV} following refs.~\cite{Christ:2015pwa,Briceno:2019opb}, the strategy to convert the finite-volume estimator to the physical observable (and thereby cancel the poles and include the imaginary contribution) is known and can be applied in this case.

The remainder of this paper is organized as follows: In section \ref{sec:pheno} we discuss the currently available phenomenological strategy and predictions to compute the $\sigmatopll$ branching ratio. Section \ref{sec:amplitude} then outlines our strategy on the lattice, which aims to recover the $\sigmatopll$ amplitude via carefully chosen, numerically calculable Euclidean correlation functions. Here we discuss various strategies to remove the exponentially growing terms that will appear in the direct lattice result. In section \ref{sec:FV} we discuss the removal of finite-volume singularities and make contact with the physical observable. In addition to translating the general formalism of refs.~\cite{Christ:2015pwa,Briceno:2019opb} to our particular case, we also provide an explicit expression for an expansion that arises when the volume is tuned so that the mass of the $\Sigma^+$ coincides with one of the finite-volume $N \pi$ energies. We close with a brief summary in section \ref{sec:summary}. This work also contains five appendices detailing various technical aspects useful for the practical calculation.



\section{Phenomenological background}
\label{sec:pheno}
\begin{figure}[h]
 \centering
 \includegraphics[width=0.85\textwidth]{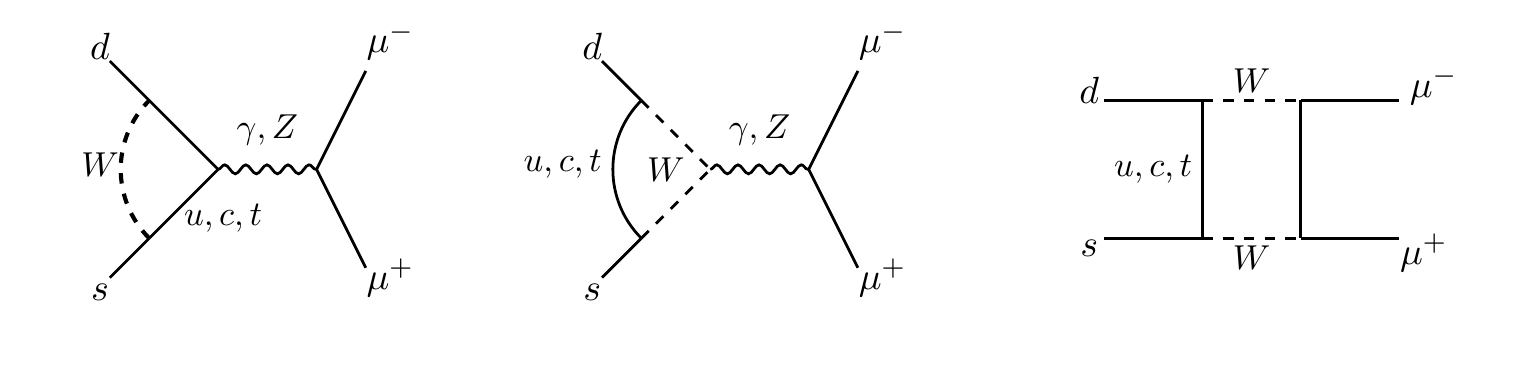}
 \vspace{-0.5cm}
 \caption{Short-distance Standard Model contributions to the $s\to d$ transition from penguin and 
box diagrams.}
 \label{fig:shortdist}
\end{figure}

In the following, we will briefly review the phenomenological determination \cite{He:2005yn,He:2018yzu,Geng:2021fog} of the 
branching ratio for $\sigmatopll$ leading to the result quoted in eq.~\eqref{eq:BRHeetal}. Short distance contributions to $\sigmatopll$ originate from penguin and box diagrams (cf. 
figure~\ref{fig:shortdist}) and are found to contribute only at the order of $10^{-12}$ 
\cite{He:2005yn} to the branching ratio of the muonic mode $\sigmatopmumu$, which is much smaller than the 
experimental measurements (eqs.~\eqref{eq:BRLHCb} and \eqref{eq:BRHyperCP}), indicating that 
this decay is long-distance dominated.
\par

The matrix element for the long-distance Standard Model contribution to the $\sigmatopll$ decay 
can be written as \cite{Bergstrom:1987wr,He:2005yn} 
\begin{equation}
    \begin{split}
    \mathcal{A}(\sigmatopll) & = - e^2G_F \times  \overline{u}_{\ell}(\boldsymbol p_{\ell^-})\,\gamma^\nu\,v_{\ell}(\boldsymbol p_{\ell^+}) \\
 &   \times  \overline{u}_p(\boldsymbol p) \bigg [
    \frac{i}{q^2} \left(a(q^2)+b(q^2)\gamma_5\right)\sigma_{\mu\nu}q^\mu
   +  \gamma_\nu \left(c(q^2)+d(q^2)\gamma_5\right)
     \bigg ]  u_{\Sigma}(\boldsymbol k)  \,,
    \end{split} 
     \label{eq:matrixelem}
    \end{equation}
    in terms of the four form factors $a(q^2)$, $b(q^2)$, $c(q^2)$ and $d(q^2)$. Here $\gamma^\nu$, $\gamma_5$ and $\sigma_{\mu \nu}$ are (combinations of) Minkowski gamma matrices, with conventions defined in appendix \ref{sec:appMinEuc}, and $u_{\Sigma}(\boldsymbol k)$, $\overline{u}_p(\boldsymbol p)$, $\overline{u}_{\ell}(\boldsymbol p_{\ell^-})$, and $v_{\ell}(\boldsymbol p_{\ell^+})$ are the usual Dirac spinors for the incoming $\Sigma^+$ and the outgoing proton, $\ell^-$, and $\ell^+$, respectively. 
    The four-momentum transfer is
    $q= k - p$, where $k$ and $p$ are the on-shell four-momenta of the $\Sigma^+$ and proton, 
    respectively.

Some information about the form factors $a$ and $b$ can be obtained from the decay $\Sigma^+\to 
p\gamma$ with a real photon. The respective decay rate\footnote{ 
The Particle Data Group \cite{Tanabashi:2018oca} quotes 
$\mathcal{B}\left(\Sigma^+\to p\gamma\right) = 1.23\pm0.05\times 
10^{-3}\quad\textrm{with}\quad \tau_{\Sigma^+} = 0.8018\pm0.0026\times 10^{-10} \  \text{s}$, 
giving $\Gamma(\Sigma^+\to p \gamma) = (10.1\pm0.4)\times 10^{-15}~\textrm{MeV}$ and $\alpha = -0.76\pm0.08$.
}
can be written as
\begin{align}
 &\Gamma(\Sigma^+\to p \gamma) =\frac{G_F^2e^2}{\pi}\,|\boldsymbol{q}|^3\,\left(|a(0)|^2+|b(0)|^2\right) \,,
 \label{eq:realphoton1}\\[0.3cm]
 &\frac{\intd\Gamma(\Sigma^+\to p \gamma)}{\intd\cos\theta}\,\, \propto\,\,  
1+\alpha\cos\theta\qquad\text{with}\qquad \alpha = 
\frac{2\,\textrm{Re}\!\left(a(0)b(0)^*\right)}{|a(0)|^2+|b(0)|^2} \,,
\label{eq:realphoton2}
\end{align}
where $|\boldsymbol{q}|$ is the energy of the photon and $\theta$ is the angle between the spin of the 
$\Sigma^+$ and the momentum of the proton.
\par
The imaginary parts of the four form factors can be obtained from unitarity using amplitudes for $\Sigma\rightarrow N\pi$ and $N\pi\rightarrow N\gamma^*$. While the amplitude for $\Sigma\rightarrow N\pi$ is known from experimental measurements \cite{Tanabashi:2018oca}, the authors of ref.~\cite{He:2005yn} calculate the amplitude $N\pi\rightarrow N\gamma^*$ from Chiral Perturbation Theory (ChiPT) using either the relativistic baryon ChiPT \cite{Bijnens:1985kj} or the heavy-baryon ChiPT \cite{Jenkins:1990jv,Jenkins:1992ab} formulation. The momentum dependence of the imaginary parts of the form factors is found to be very mild. Once the imaginary parts of the form factors are known, information on the real parts of $a(q^2)$ and $b(q^2)$ at $q^2=0$ can be obtained from equations \eqref{eq:realphoton1} and \eqref{eq:realphoton2} and experimental data for the decay $\Sigma^+\to p \gamma$. Since this decay only determines values for $|a(0)|^2+|b(0)|^2$ and $\textrm{Re}\!\left[a(0)b(0)^*\right]$, this leads to four possible solutions for $\big(\textrm{Re} \big[a(0)\big]$, $\textrm{Re} \big[b(0)\big]\big)$. Motivated by the mild $q^2$-dependence of the imaginary parts of the form factors, the authors of ref.~\cite{He:2005yn} assume that  
$\textrm{Re}\big[a(q^2)\big]= \textrm{Re} \big[a(0)\big]$ and $\textrm{Re}\big[b(q^2)\big]= \textrm{Re} \big[b(0)\big]$ for their prediction of $\Bsigmatopll$. The real parts of the $c(q^2)$ and $d(q^2)$ form factors are calculated assuming vector meson dominance in \cite{He:2005yn} and explicitly calculating vector meson pole contributions to the decay $\sigmatopll$. The $q^2$-dependence of $\textrm{Re}\big[c(q^2)\big]$ and $\textrm{Re}\big[d(q^2)\big]$ is found to be mild, just like in the imaginary parts.
\par
Depending on the formulation of baryon ChiPT used and the four possible solutions for $\textrm{Re} \big[a(0)\big]$ and $\textrm{Re} \big[b(0)\big]$ from $\Sigma^+\to p \gamma$ decays, the authors of ref.~\cite{He:2005yn} find the Standard Model prediction for $\Bsigmatopmumu$ to be in the range
\begin{equation}
 1.6\times 10^{-8}\leq \mathcal{B}\left(\sigmatopmumu\right) \leq 9.0\times10^{-8}\,,
\label{eq:Heetal}
 \end{equation}
  and very similar ranges are found in refs.~\cite{He:2018yzu,Geng:2021fog}.



\section{Extracting the amplitude from Euclidean correlators}
\label{sec:amplitude}

In this section and the next, we describe how to extract the $\sigmatopll$ amplitude from a numerical lattice calculation. The approach closely follows the methods of refs.~\cite{\AllLongRange}, adjusted here to treat issues specific to this system. This section details the Euclidean two-, three-, and four-point correlation functions needed to construct a finite-volume estimator, denoted by $\widetilde F_\mu(\boldsymbol k, \boldsymbol p)_L$. The following section describes how to relate this quantity to the physical rare hyperon amplitude.

\subsection{Spectral representation}

The determination of the long-distance contribution to $\sigmatopll$ requires a calculation of the $\Sigma^+\rightarrow p\gamma^*$ amplitude, defined as
\begin{equation}
\mathcal{A}_\mu^{rs}(k, p) =
\int\intd^4x\,\big<p(\boldsymbol{p}),r\big|\textrm{T}\left[\mathcal H_W(x)\,J_\mu(0)\right]\big|\Sigma^+(\boldsymbol{
k}),s\big> \,,
\label{eq:hadamplitude}
\end{equation}
with $r$ and $s$ labelling the azimuthal spin component of the state. Here we are assuming Minkowski-signature conventions and working in an infinite space-time volume. This amplitude can be re-expressed as a Dirac matrix, $\widetilde{\mathcal{A}}_\mu(k, p)$, using the relation
\begin{equation}
\label{eq:spinmatrix}
\mathcal{A}_\mu^{rs}(k, p) = \overline{u}_p^r(\boldsymbol p) \,\widetilde{\mathcal{A}}_\mu(k, p)\,u^s_{\Sigma}(\boldsymbol k) \,,
\end{equation}
with the spinors $u_p$ and $u_{\Sigma}$ of the proton and $\Sigma^+$, respectively.\footnote{To avoid clutter in notation, we denote indices for the $\Sigma^+$ by $\Sigma$ only. We still use $\Sigma^+$ where the charge is relevant, e.g. for creation and annihilation operators.} 

The effective weak-Hamiltonian density of the $qs\rightarrow qd$
transition is given by \cite{Buchalla:1995vs}
\begin{equation}
\mathcal H_W(x) = \frac{G_F}{\sqrt{2}} V_{us}V^*_{ud}\big[C_1 \big (Q_1^{u}(x)-Q_1^{c}(x) \big )+C_2 \big ( Q_2^{u}(x)-Q_2^{c}(x) \big ) + \cdots \big] \,,
\label{eq:weakH}
\end{equation}
where the $C_{i}$ are Wilson coefficients, the $Q_i^q$ are four-quark operators, defined in terms of Dirac spinors for up, down, strange and charm quarks (respectively $u,d,s$ and $c$) as
\begin{equation}
Q^q_1 =
(\overline{d}_{i}\gamma^{L\,\mu}s_{i})(\overline{q}_{j}\gamma^L_{
\mu}q_{j}) \,, \hspace{1cm} Q^q_2 =
(\overline{d}_{i}\gamma^{L\,\mu}q_{i})(\overline{q}_{j}\gamma^L_{\mu}s_{j}) \,,
\label{eq:4QuarkOp}
\end{equation}
and it is understood that the two are renormalized in the same scheme, at the same scale, such that the weak Hamiltonian is scheme independent. Here $i,j$ denote colour indices and we define $\gamma^L_{\mu}\equiv\gamma_\mu(1-\gamma_5)$. There are additional four-quark operators in \eqref{eq:weakH} with Wilson coefficients of order $|\frac{V_{ts} V_{td}}{V_{us} V_{ud}}| \simeq 0.00142$ which will be neglected in this work. 

The
electromagnetic current in eq.~\eqref{eq:hadamplitude} is given by
\begin{equation}
J_\mu = \frac{2}{3}\overline{u}\gamma_\mu u - \frac{1}{3}\overline{d}\gamma_\mu d
- \frac{1}{3}\overline{s}\gamma_\mu s + \frac{2}{3}\overline{c}\gamma_\mu c\,,
\end{equation}
and we make use of translational invariance by fixing the position of the electromagnetic current to $y = 0$. Including an additional Fourier transform on the current would lead to an overall momentum-conserving Dirac delta function, to be removed at a later step, and we find it more convenient to follow the approach where this is never introduced.

The Hamiltonian density decomposes into a parity-positive and a parity-negative component
\begin{equation}
\mathcal H_W(x) = \mathcal H^+_W(x) + \mathcal H^-_W(x) \,,
\end{equation}
defined via the parity operator $\hat{\mathcal{P}}$ according to
\begin{equation}
\hat{\mathcal P} \mathcal H^\pm_W(x) \hat{\mathcal P} = \pm \mathcal H^\pm_W(\mathcal P \cdot x) \,,
\end{equation}
where ${\mathcal P^{\mu}}_\nu = \text{diag}\big [1, -1, -1, -1 \big]$ and $\hat{\mathcal P}$ is the Hilbert-space representation of the parity operator. Both parity sectors contribute to the amplitude we are evaluating.

Defining $ \widetilde {\mathcal A}_\mu^{\pm}(k, p)$ as in eqs.~\eqref{eq:hadamplitude} and \eqref{eq:spinmatrix}, but with $ \mathcal H_W(x) $ replaced by $ \mathcal H^\pm_W(x) $, one can next decompose each definite-parity amplitude in terms of form factors as follows \cite{Bergstrom:1987wr,He:2005yn}:
\begin{align}
\widetilde{\mathcal{A}}^+_\mu(k,p) & = i\sigma_{\nu\mu}q^\nu a(q^2) +
\left(q^2\gamma_\mu-q_\mu \slashed{q}\right) c(q^2) \,, \\
\widetilde{\mathcal{A}}^-_\mu(k,p) & = i\sigma_{\nu\mu}q^\nu \gamma_5 b(q^2) +
\left(q^2\gamma_\mu-q_\mu \slashed{q}\right) \gamma_5 d(q^2) \,,
\end{align}
where we recall that $q=k-p$ is the four-momentum transfer of the virtual photon. 
This form-factor decomposition is derived in Appendix \ref{sec:appFFdec}.
Note also that, while the amplitude is a Dirac matrix and thus depends on individual components of the four-momenta, the form factors are Lorentz scalars and can therefore only depend on $q^2$.

We will see in the following that the amplitude, and thus also the form factors, are complex-valued due to on-shell intermediate $N \pi$ states. Since the Euclidean correlators are real-valued, this complexity already signals the fact that it is non-trivial to extract the amplitudes. This turns out to be closely related to the interplay of the Euclidean signature and the finite volume. As we will show in the following, the quantum numbers of the contributing $N \pi$ states differ for $\widetilde{\mathcal{A}}^+_\mu$ and $\widetilde{\mathcal{A}}^-_\mu$, and thus the finite-volume formalism must be applied independently to the two quantities.

To explain this in more detail we return to eq.~\eqref{eq:hadamplitude} and insert a complete set of states between the current and the weak Hamiltonian to write
\begin{equation}
\widetilde {\mathcal{A}}_\mu^{\pm}(k,p) =
\int_0^\infty \intd \omega \, \bigg [ \int_{-\infty}^0 \intd t \, \widetilde{\rho}^{\pm }_{\mu}(\omega)\, e^{- i ( E_{\Sigma}(\boldsymbol k) -\omega + i \epsilon ) t} + \int_{0}^\infty \intd t \, \widetilde{\sigma}^{\pm }_{\mu}(\omega)\, e^{- i ( \omega - E_{p}(\boldsymbol p) - i \epsilon ) t} \bigg ] \,,
\end{equation}
where we have introduced the spectral functions, satisfying
\begin{align}
\label{eq:rhodef}
\!\!\!  \overline{u}_p^r(\boldsymbol p) \,  \widetilde{\rho}^{\pm}_{\mu}(\omega)  \,u^s_{\Sigma}(\boldsymbol k)   & = \sum_{\alpha }  \frac{\delta(\omega - E_{\alpha}(\boldsymbol k))}{2 E_{\alpha}(\boldsymbol k)}   \,
\big < p(\boldsymbol{p}),r \big| J_\mu(0) \big | E_{\alpha}, \boldsymbol k \big> \big < E_{\alpha}, \boldsymbol k \big | {\mathcal H^{\pm}_{W}}(0) \big|\Sigma^+(\boldsymbol{k}),s\big>  \,, \\
\label{eq:sigmadef}
\!\!\!  \overline{u}_p^r(\boldsymbol p) \,  \widetilde{\sigma}^{\pm}_{\mu}(\omega)  \,u^s_{\Sigma}(\boldsymbol k)  & = \sum_{\beta} \frac{\delta(\omega - E_{\beta}(\boldsymbol p))}{2 E_{\beta}(\boldsymbol p)}  \, \big < p(\boldsymbol{p}),r \big| {\mathcal H^{\pm}_{W}}(0) \big | E_{\beta}, \boldsymbol p \big> \big < E_{\beta}, \boldsymbol p \big | J_\mu(0) \big|\Sigma^+(\boldsymbol{k}),s\big> \,.
\end{align}
Note that one must treat the two time orderings separately and this leads to two types of intermediate states encoded in $\widetilde{\rho}$ and $\widetilde{\sigma}$, which have strangeness $S=0$ and $S=-1$ respectively. The sums over $\alpha$ and $\beta$ represent both sums and phase-space integrals over the multi-particle QCD Fock space for all states that contribute. For example, the sum over $\alpha$ includes $N \pi$, $N \pi \pi$, $\Delta \pi$ and $\Lambda K$ states.

Evaluating the time integrals then gives a compact result
\begin{equation}
\widetilde {\mathcal{A}}_\mu^{\pm}(k,p) = i
\int_0^\infty \intd \omega \, \frac{\widetilde{\rho}^{\pm}_{\mu}(\omega)}{ E_{\Sigma}(\boldsymbol k) -\omega + i \epsilon } - i \int_0^\infty \intd \omega \, \frac{\widetilde{\sigma}^{\pm}_{\mu}(\omega)}{ \omega - E_{p}(\boldsymbol p) - i \epsilon} \,.
\label{eq:Minkowskiamplitude}
\end{equation}
The aim of the following sections is to review how this amplitude can be extracted from finite-volume Euclidean-signature correlation functions.

\subsection{Euclidean correlation functions}

We now discuss how to extract a finite-volume estimator for the desired Minkowski-space amplitude \eqref{eq:Minkowskiamplitude} from Euclidean correlation functions that can be calculated on the lattice. All quantities in this section (e.g.~Dirac $\gamma$-matrices, four-vectors) are defined with Euclidean conventions detailed in Appendix \ref{sec:appMinEuc}.

\subsubsection{Two-point functions}
The two-point function of a baryon $B$ can be written as
\begin{equation}
\Gamma^{(2)}_B(t,\boldsymbol{p})_{\alpha\beta} = \int \intd^3\boldsymbol{x} \, e^{-i \boldsymbol p \cdot \boldsymbol x} \left<\psi^B_\alpha(t,\boldsymbol{x}) \,\overline{\psi}
^B_\beta(0)\right>\,,
\label{eq:twopt}
\end{equation}
where $\overline{\psi}^B_\beta(t,\boldsymbol{x})$ and $\psi^B_\alpha(t,\boldsymbol{x})$ create and annihilate, respectively, a baryon $B$ and $\alpha$ and $\beta$ are Dirac spinor indices.
Examples for operators that have overlap with the proton $p$ and $\Sigma^+$ are
\begin{align}
  \psi^p_\delta(t,\boldsymbol{x}) = \epsilon^{abc}\,(P^+\Gamma^A)_{\delta\gamma}\,u_{c,\gamma}(x)\,\left(u_{a,\alpha}(x)\,\Gamma^B_{\alpha\beta}\,d_{b,\beta}(x)\right) \,, \\
  \psi^{\Sigma^+}_\delta(t,\boldsymbol{x}) =  \epsilon^{abc}\,(P^+\Gamma^A)_{\delta\gamma}\,u_{c,\gamma}(x)\,\left(u_{a,\alpha}(x)\,\Gamma^B_{\alpha\beta}\,s_{b,\beta}(x)\right) \,,
  \end{align}
with $\Gamma^A=\mathds{1}$, $\Gamma^B=C\gamma_5$ for spin $1/2$ particles and the charge conjugation operator defined by $C=\gamma_0\gamma_2$. $P^+ = \frac{1}{2}(1+\gamma_0)$ projects to the positive parity state. Here, and below, Roman and Greek indices refer to colour and Dirac indices, respectively.

For $t>0$, the two-point function \eqref{eq:twopt} has the spectral representation
\begin{equation}
\Gamma^{(2)}_B(t,\boldsymbol{p}) = \sum_s \frac{|Z_B|^2 u_B^s \overline{u}_B^s}{2\,E_B(\boldsymbol{p})}e^{-t E_B(\boldsymbol{p})} =
|Z_B|^2 \, \frac{M_B}{E_B(\boldsymbol p)} \, \mathbb P_B(\boldsymbol p) e^{-t E_B(\boldsymbol{p})} \,,
\label{eq:specreptwopoint}
\end{equation}
plus contributions from excited states, which are exponentially suppressed in $t$ by their higher energies. The moving-frame energy of the baryon is given by $E_B(\boldsymbol{p})=\sqrt{M_B^2+\boldsymbol{p}^2}$ and the four-momentum vector is $p_B \equiv (iE_B(\boldsymbol{p}),\boldsymbol{p})$. We have introduced the projector
\begin{equation}
\mathbb P_B(\boldsymbol p) = \frac{1}{2 M_B} \sum_r  u^r_B(\boldsymbol p) \bar u^r_B(\boldsymbol p) = \frac{ (-i\slashed{p}_B+M_B)}{2 M_B} \,.
\end{equation} 
The overlap factor $Z_B$ is defined by
\begin{equation}
\big<B(\boldsymbol{p}),s\big|  \overline{\psi}^B(0)   \big|0\big>_{\infty} = Z_B^\dagger(\boldsymbol{p})\,\,\overline{u}^s_B(\boldsymbol{p}_B) \,,
\hspace{1cm}
\big<0\big|   \psi^B(0)   \big|B(\boldsymbol{p}),s\big>_{\infty} = Z_B(\boldsymbol{p})\,\,u^s_B(\boldsymbol{p}_B) \,.
\end{equation}
Throughout this section and the next, we neglect the finite-volume effects on single-hadron energies and matrix elements. These are known to be exponentially suppressed, scaling as $e^{- M_\pi L}$.
We also include the $\infty$ label on states that are normalized according to 
\begin{equation}
\big < B(\boldsymbol{p'}),s' \big \vert B(\boldsymbol{p}),s \big >_{\infty} = 2 E_B(\boldsymbol p) \delta_{s's} (2 \pi)^3 \delta^3(\boldsymbol p' - \boldsymbol p) \,,
\end{equation}
as this differs from the normalization of finite-volume states used below.

\subsubsection{Three-point functions}
\label{subsubsec:threept}
We turn now to the three-point function of the effective weak Hamiltonian $\mathcal H_W$ between a $\Sigma^+$ and a $p$ operator
\begin{equation}
\Gamma^{(3)}_H(t_H,t_p,t_{\Sigma};\boldsymbol{p})_{\alpha\beta} =
\int\intd^3 \boldsymbol{x}_p \intd^3 \boldsymbol{x}_{\Sigma}\,e^{- i \boldsymbol p \cdot ( \boldsymbol x_p - \boldsymbol x_{\Sigma})}\,
\left<\psi^p_\alpha(t_p,\boldsymbol{x}_p)\,\,{\mathcal H_{W}}(t_H,\boldsymbol{0})\,\,\overline{\psi}
^{\Sigma^+}_\beta(t_{\Sigma},\boldsymbol{x}_{\Sigma}) \right> \,,
\label{eq:threeptH}
\end{equation}
with the effective weak-Hamiltonian density ${\mathcal H_{W}}$ given in eq.~\eqref{eq:weakH}. We leave the parity labels implicit throughout this section unless stated explicitly.

Here the weak-Hamiltonian is assumed to be appropriately renormalised. In the case of chirally symmetric fermion discretisations and a massless renormalisation scheme, the 4-quark operators in \eqref{eq:4QuarkOp} are protected from mixing with other dimension-6 operators. In fact, the $V-A$ structure of our particular 4-quark operator guarantees that all divergences are removed with massless renormalisation factors, even when the matrix elements themselves are evaluated at non-zero quark mass. This is described in detail, in the context of domain-wall-fermion lattice calculations, in refs. \cite{Christ:2012se,Bai:2014cva}, where the $K_L - K_S$ mass difference is evaluated using the same operator. The key point for the present work is that no aspects of the renormalisation are affected by the external states, so we can directly adopt the strategy of the earlier publications. See also the text in the paragraph after eq.~\eqref{eq:fourptHJ}.

\begin{figure}[h]
\centering
\includegraphics[width=0.9\textwidth]{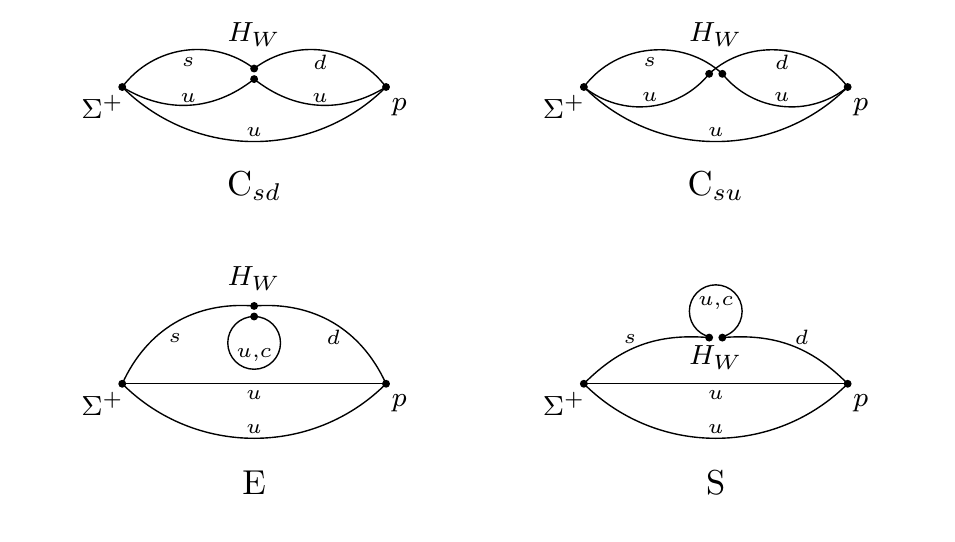}
\caption{The four different topologies for the Wick contractions of the three-point function
\eqref{eq:threeptH}. Two fully connected contractions C$_{sd}$ and
C$_{su}$ and two topologies (E and S) containing quark loops. The two diagrams on the left (C$_{sd}$
and E) arise from contractions using the $Q_1$ operator, the two diagrams on the right (C$_{su}$
and S) from $Q_2$.}
\label{fig:threeptdiagrams}
\end{figure}

The three-point function \eqref{eq:threeptH} gives rise to four different topologies\footnote{Naming conventions for the diagrams shown in Figure~\ref{fig:threeptdiagrams} are inspired by ref.~\cite{Christ:2015aha}.} for the Wick contractions, which are shown in Figure \ref{fig:threeptdiagrams}. The double point labelled with ${\mathcal H_{W}}$ shows the position of the weak Hamiltonian, and the points labelled $\Sigma^+$ and $p$ are the positions of the ${\Sigma^+}$ and proton operator, respectively. The quark lines are labelled by their respective quark flavours. The two diagrams shown on the left-hand side of Figure \ref{fig:threeptdiagrams} arise from contractions using the $Q_1$ operator in the weak Hamiltonian, the two diagrams on the right-hand side arise from contractions using the $Q_2$ operator.

The spectral representation of the Euclidean three-point function $\Gamma^{(3)}$ is given by
\begin{equation}
\Gamma^{(3)}_H(t_H,t_p,t_{\Sigma};\boldsymbol{p}) =\sum_{r,s} \frac{Z_p(\boldsymbol{p})\,Z^\dagger_{\Sigma}(\boldsymbol{p})\,\,u^r_p(\boldsymbol{p})\,\,\mathcal{A}_H^{rs}\,\,\overline{u}^s_{\Sigma}(\boldsymbol{p})}{4E_p(\boldsymbol{p})E_{\Sigma}(\boldsymbol{p})}\,e^{-E_p(\boldsymbol{p})\,(t_p-t_H)}\,e^{-E_{\Sigma}(\boldsymbol{p})\,(t_H-t_{\Sigma})} \,,
\label{eq:specrepthreept}
\end{equation}
for large time-separations $t_{\Sigma}\ll t_H$ and $t_H\ll t_p$, such that excited states are suppressed, with
\begin{equation}
\mathcal{A}_H^{rs} = \big \langle p(\boldsymbol{p}),r \big \vert {\mathcal H_{W}}(0) \big \vert {\Sigma^+}(\boldsymbol{p}),s \big \rangle_\infty \,\,\equiv\,\, \overline{u}^r_p(\boldsymbol{p})\,\,\widetilde{\mathcal{A}}_H\,\,u^s_{\Sigma}(\boldsymbol{p})\,,
\end{equation}
where $r$ and $s$ are the spins of the proton and $\Sigma^+$, respectively. In the following it will be convenient to define the overall normalization factor
\begin{equation}
Z_{BB'}(t_B,t_{B'};\boldsymbol{p},\boldsymbol{k}) \equiv \frac{Z_B(\boldsymbol{p})\,Z^\dagger_{B'}(\boldsymbol{k})M_B M_{B'}}{E_B(\boldsymbol{p})E_{B'}(\boldsymbol{k})} e^{-E_B(\boldsymbol{p})\,t_B}\,e^{E_{B'}(\boldsymbol{k})\,t_{B'}}\,,
\label{eq:ZNSigma}
\end{equation}
where $B,B' \in \{ p, \Sigma \}$. This factor can be constructed using information extracted from $\Sigma^+$ and proton two-point functions (cf.~eq.~\eqref{eq:specreptwopoint}). Completing the spin sum, the spectral representation \eqref{eq:specrepthreept} can be written as
\begin{equation}
\Gamma^{(3)}_H(t_H,t_p,t_{\Sigma};\boldsymbol{p}) = Z_{p{\Sigma}}(t_p,t_{\Sigma};\boldsymbol{p},\boldsymbol{p}) \mathbb P_p(\boldsymbol p) \, \widetilde{\mathcal{A}}_H \,  \mathbb P_{\Sigma}(\boldsymbol p)\,e^{-t_H [ (E_{\Sigma}(\boldsymbol{p}) - E_p(\boldsymbol{p}) ]}\,,
\label{eq:3ptH}
\end{equation}
plus contributions from excited states.
Similarly, one can define the three-point function for a baryon $B$ (here $B \in \{p,\Sigma\}$) with an electromagnetic current $J_\mu$
\begin{equation}
\Gamma^{(3)}_{\mu,B}(t_J,t_f,t_i;\boldsymbol{p},\boldsymbol{k})_{\alpha\beta} =
\int\intd^3 \boldsymbol{x}_f \intd^3 \boldsymbol{x}_i\,e^{- i (\boldsymbol p \cdot \boldsymbol x_f - \boldsymbol k \cdot \boldsymbol x_i)}\,
\left<\psi^B_\alpha(t_f,\boldsymbol{x}_f)\,\,J_\mu(t_J,\boldsymbol{0})\,\,\overline{\psi}
^B_\beta(t_i,\boldsymbol{x}_i) \right>\,.
\label{eq:threeptJ}
\end{equation}
The spectral representation of such a three-point function is given by
\begin{align}
\Gamma^{(3)}_{\mu,B}(t_J,t_f,t_i;\boldsymbol{p},\boldsymbol{k})
&= Z_{BB}(t_f,t_i;\boldsymbol{p},\boldsymbol{k}) \mathbb P_B(\boldsymbol p) \, \widetilde{\mathcal{A}}_{\mu,B}(\boldsymbol{q})\, \mathbb P_B(\boldsymbol k)   e^{-t_J (E_B(\boldsymbol{k}) - E_B(\boldsymbol{p})) } \,,
\label{eq:3ptJ}
\end{align}
with   $\widetilde{\mathcal{A}}_{\mu,B}$ defined by
\begin{equation}
\mathcal{A}_{\mu,B}^{rs}(\boldsymbol{q}) = \big<B(\boldsymbol{p}),r\big| \,J_\mu(0) \,\big|B(\boldsymbol{k}),s\big>_\infty   \equiv  \overline{u}^r_B(\boldsymbol{p})\,\,\widetilde{\mathcal{A}}_{\mu,B}(\boldsymbol{q})\,\,u^s_B(\boldsymbol{k})\,.
\end{equation}
Just as with the weak-Hamiltonian operator, the electromagnetic vector current is also assumed to be renormalised. If the conserved Noether current is used, then the vector Ward identity is exactly obeyed on the lattice and no renormalisation is necessary.

Here we have only considered three-point functions with single-hadron states. However, as discussed in the following subsection, to construct the target finite-volume estimator, one may also require matrix elements involving the finite-volume analogue of a multi-particle excited state. Many details of the construction of $\mathcal{A}_H^{rs} $ and $\mathcal{A}_{\mu,B}^{rs}(\boldsymbol{q})$ also apply for the excited-state analogues, but important differences arise. We describe this in detail in Appendix~\ref{app:details_fvstates}.

Finally, we define amputated versions of three-point functions with certain factors removed. For the vector current we take
\begin{equation}
\begin{split}
\hat{\Gamma}^{(3)}_{\mu,p}(t_J; \boldsymbol{p},\boldsymbol{k}) &\equiv
\frac{\Gamma^{(3)}_{\mu,p}(t_J,t_f,t_i;\boldsymbol{p},\boldsymbol{k})}{Z_{pp}(t_i,t_f;\boldsymbol{p},\boldsymbol{k})} =   \mathbb P_p(\boldsymbol p) \,\big[ \widetilde{\mathcal{A}}_{\mu,p}(\boldsymbol{q}) e^{- t_J (E_{p}(\boldsymbol k) - E_p(\boldsymbol p))} \big ]\, \mathbb P_p(\boldsymbol k) \,,
\end{split} \label{eq:three_vec_hat}
\end{equation}
and for the weak Hamiltonian
\begin{equation}
\hat{\Gamma}^{(3)}_H(t_H;\boldsymbol{k})\equiv \frac{\Gamma^{(3)}_H(t_H,t_p,t_{\Sigma};\boldsymbol{k})}{Z_{p{\Sigma}}(t_p,t_{\Sigma};\boldsymbol{k},\boldsymbol{k})} =   \mathbb P_p(\boldsymbol k) \, \big [ \widetilde{\mathcal{A}}_H \, e^{- t_H (E_{\Sigma}(\boldsymbol k) - E_p(\boldsymbol k))} \big ] \, \mathbb P_{\Sigma}(\boldsymbol k)\,.
\label{eq:three_hamil_hat}
\end{equation}
In both cases we drop time dependence which cancels in the limit that the ground-state dominates.

\subsubsection{Four-point functions}
We turn now to the four-point function of the time-ordered product of the weak-Hamiltonian density, ${\mathcal H_{W}}(x)$, and the electromagnetic current, $J_\mu(0)$, between a ${\Sigma^+}$ and a $p$ state
\begin{multline}
\Gamma^{(4)}_{\mu, \alpha\beta}(t_H,t_p,t_{\Sigma};\boldsymbol{p},\boldsymbol{k})_{L,T} =
\int_L \intd^3 \boldsymbol{x}\intd^3 \boldsymbol{x}_p\intd^3 \boldsymbol{x}_{\Sigma} \, e^{- i (\boldsymbol x_p \cdot \boldsymbol p - \boldsymbol x_{\Sigma} \cdot \boldsymbol k)} \\ \times
\left<\psi^p_\alpha(t_p,\boldsymbol{x}_p)\,\,\textrm{T}\left[{\mathcal H_{W}}(t_H,\boldsymbol{x})\,J_\mu(0)\right]\,\,\overline{\psi}
^{\Sigma^+}_\beta(t_{\Sigma},\boldsymbol{x}_{\Sigma}) \right>_{L,T} \,.
\label{eq:fourptHJ}
\end{multline}
Here the subscripts $L,T$ indicate that the quantity is evaluated in a finite space-time volume. This is of particular importance for the four-point function, so we emphasize the fact with our notation. Note also that the lattice path integral will always give the time-ordered product of all four-fields. We restrict attention to the case of $t_{\Sigma} < 0,t_H < t_p$ so that the fields can be written as shown.

The weak-Hamiltonian and electromagnetic current operators should be renormalised in the same way as the relevant three-point functions above to remove divergences coming from the operators themselves. However, additional divergences can come from the contact of the two operators $x=(t_H, \vect{x}) \to 0$. Refs. \cite{Isidori:2005tv,Christ:2015aha} describe in detail the origin and cancellation of these divergences, which we summarise here. From power counting, the contact term between $\mathcal{H}_W$ and $J_\mu$ can diverge at most quadratically. However, use of the conserved Noether current for the electromagnetic operator gives exact QED gauge invariance, which reduces the degree of divergence by two down to a logarithmic one. Finally, this logarithmic divergence is independent of the quark mass, and therefore cancels in the GIM subtraction between the up and charm quarks $Q_i^u-Q_i^c$ in \eqref{eq:weakH}. Therefore, so long as a conserved electromagnetic vector current is used, there are no divergences as $\mathcal{H}_W$ approaches $J_\mu$. While this previous work has been performed in the context of the rare Kaon decay $K \to \pi \ell^+ \ell^-$, since none of the arguments rely on the nature of the external states, no additional modifications are required for use in the baryonic decay $\Sigma^+ \to p \ell^+ \ell^-$.

As with the Minkowski amplitude and the three-point function considered above, $\boldsymbol{q}=\boldsymbol{k}-\boldsymbol{p}$ denotes the momentum transfer at the electromagnetic vertex. From the four-point function, one obtains six Wick contractions for each of the four topologies (cf.~Figure~\ref{fig:threeptdiagrams}) of the three-point function with the weak Hamiltonian: the electromagnetic current $J_\mu$ can be inserted on any of the five quark lines or on a disconnected quark-loop. The diagrams corresponding to the in total 24 Wick contractions are shown in Appendix \ref{sec:appWick}.

We next remove the overlap factor $Z_{p{\Sigma}}(t_p,t_{\Sigma};\boldsymbol{p},\boldsymbol{k})$, given in eq.~\eqref{eq:ZNSigma}, to define $\hat{\Gamma}^{(4)}_\mu$ as
\begin{equation}
\label{eq:gammahatdef}
\hat{\Gamma}^{(4)}_\mu(t_H;\boldsymbol{p},\boldsymbol{k})_{L} = Z_{p{\Sigma}}(t_p,t_{\Sigma};\boldsymbol{p},\boldsymbol{k})^{-1} \, \Gamma^{(4)}_\mu(t_H,t_p,t_{\Sigma};\boldsymbol{p},\boldsymbol{k})_{L,T} \,.
\end{equation}
We drop the dependence on $T$, $t_p$, and $t_{\Sigma}$ in this quantity since the ratio is independent of these time coordinates as long as $ T \gg t_p$, $T \gg \vert t_{\Sigma} \vert \gg  1/\Delta E_{\Sigma}$ and  $\vert t_p \vert \gg 1/\Delta E_{p}$, where $\Delta E_{\Sigma}$, $\Delta E_{p}$ are the gaps between the ground and first-excited states for the quantum numbers indicated by the subscript. We view it as a task of the numerical analysis to remove or quantify residual dependence on $ T, t_p,$ and $ \vert t_{\Sigma} \vert$ and omit these coordinates for the remainder of this work. 

The amputated four-point function is then equal to the matrix element
\begin{equation}
  \hat{\Gamma}^{(4)}_\mu(t_H;\boldsymbol{p},\boldsymbol{k})_{L}  =   \mathbb P_p(\boldsymbol k) \, \widetilde {\mathcal A}_{\mu}(t_H; \boldsymbol{p}, \boldsymbol{k})_L \, \mathbb P_{\Sigma}(\boldsymbol k)\,,
  \end{equation}
where $\widetilde {\mathcal A}_{\mu}(t_H; \boldsymbol{p}, \boldsymbol{k})_L$ is defined implicitly through
\begin{align}
  \mathcal A^{rs}_{\mu}(t_H; \boldsymbol p, \boldsymbol k)_L  & \equiv \overline{u}^r_p(\boldsymbol{p})\,\, \widetilde {\mathcal A}_{\mu}(t_H; \boldsymbol{p}, \boldsymbol{k})_L \,\,u^s_{\Sigma}(\boldsymbol{p}) \,,
\end{align}
where we have introduced the finite-volume, Euclidean-time-dependent analogue of eq.~\eqref{eq:hadamplitude}
\begin{align}
  \mathcal A^{rs}_{\mu}(t_H; \boldsymbol p, \boldsymbol k)_L & \equiv  
    \int_L \intd^3 \boldsymbol{x}   \, 
    \big \langle {p(\boldsymbol p), r} \big \vert  \textrm{T}\left[{\mathcal H_{W}}(t_H,\boldsymbol{x}) J_\mu(0)\right]  \big \vert {\Sigma^+( \boldsymbol k), s  } \big \rangle_{L}  \,.
  \end{align}

Inserting a complete set of finite-volume states between the current and weak-Hamiltonian density, one can give a spectral representation of the four-point function in Euclidean space-time as
\begin{subnumcases}
{\label{eq:gammahatdecom} \hat{\Gamma}^{(4)}_\mu(t_H;\boldsymbol{p},\boldsymbol{k})_{L} = }
 \int_0^\infty\intd \omega \, \mathbb P_{p}(\boldsymbol p) \, \widetilde{\sigma}_{\mu}(\omega)_{L} \, \mathbb P_{\Sigma}(\boldsymbol k) \, e^{-t_H\left[ \omega - E_p(\boldsymbol p)\right]} \,, & for $t_H > 0$ \,, \\[8pt]
 \int_0^\infty\intd \omega \, \mathbb P_{p}(\boldsymbol p) \, \widetilde{\rho}_{\mu}(\omega)_{L} \, \mathbb P_{\Sigma}(\boldsymbol k) \, e^{-t_H\left[ E_{\Sigma}(\boldsymbol k) - \omega \right]} \,, & for $t_H < 0$ \,,
\end{subnumcases}
where $\widetilde{\rho}_{\mu}(\omega)_{L}$ and $\widetilde{\sigma}_{\mu}(\omega)_{L}$ are defined as in eqs.~\eqref{eq:rhodef} and \eqref{eq:sigmadef} above but here with the Euclidean conventions in the gamma matrices and with the sum now running over the discrete finite-volume spectrum.
For example, $\widetilde \rho_\mu(\omega)_L$ can be written as
\begin{equation}
\widetilde{\rho}_{\mu}(\omega)_L = \sum_n \, \frac{ \widetilde{C}_{n, \mu}(\boldsymbol k) }{2 E_{n}(\boldsymbol k) } \, \delta \big ( E_{n}(\boldsymbol k) -\omega \big ) \,,
\end{equation}
where
\begin{equation}
\label{eq:Cdef}
\bar u^r_p(\boldsymbol p) \,  \widetilde{C}_{n, \mu}(\boldsymbol k) \, u^s_{\Sigma}(\boldsymbol k) \equiv    \big < p(\boldsymbol{p}),r \big| J_\mu(0) \big | E_{n}, \boldsymbol k \big>_L \, \big < E_{n}, \boldsymbol k \big | {\mathcal H_{W}}(0) \big|\Sigma^+(\boldsymbol{k}),s\big>_L  \,,
\end{equation}
and $\big | E_{n}, \boldsymbol k \big>_L$ is the $n$th finite-volume state with the relevant quantum numbers to contribute, normalized as $\big < E_{n}, \boldsymbol k \big | E_{n}, \boldsymbol k \big>_L = 2 E_n(\boldsymbol k)$. 

\subsection{Finite-volume estimator for the decay amplitude}

Our aim now is to extract the infinite-volume, Minkowski-signature amplitude (with spectral representation given in eq.~\eqref{eq:Minkowskiamplitude}) from the finite-volume Euclidean four-point function, decomposed above in eq.~\eqref{eq:gammahatdecom}. To do so, one needs to separately treat the issues of Euclidean time and finite volume, and we find it most instructive to address the first point in this subsection and the second in the section following. To this end we define a physical-energy finite-volume estimator of eq.~\eqref{eq:Minkowskiamplitude} as follows:
\begin{equation}
\widetilde{F}_\mu(\boldsymbol{k},\boldsymbol{p})_{L}
\equiv i
\int_0^\infty \intd \omega \, \frac{\widetilde{\rho}_{\mu}(\omega)_L}{ E_{\Sigma}(\boldsymbol k) -\omega } - i \int_0^\infty \intd \omega \, \frac{\widetilde{\sigma}_{\mu}(\omega)_L}{ \omega - E_{p}(\boldsymbol p) } \,.
\end{equation}

This definition looks similar to eq.~\eqref{eq:Minkowskiamplitude} but with the key difference that finite-volume spectral functions have been substituted. As a result, the $i \epsilon$ pole prescription has also been discarded, since this has no effect in the finite volume. To see this more explicitly we write out the first term by substituting the definition of $\widetilde{\rho}_{\mu}(\omega)_L$
\begin{equation}
\int_0^\infty \intd \omega \, \frac{\widetilde{\rho}_{\mu}(\omega)_L}{ E_{\Sigma}(\boldsymbol k) -\omega } = \sum_n \, \frac{ \widetilde{C}_{n, \mu}(\boldsymbol k) }{2 E_{n}(\boldsymbol k) \big ( E_{\Sigma}(\boldsymbol k) - E_{n}(\boldsymbol k) \big ) } \,.
\end{equation}
The sum over $n$ runs over the discrete set of finite-volume states including the proton-like ground state and multi-hadron excited states that can be related to proton-pion and other scattering amplitudes. A subtlety of this analysis is that, for non-zero $\boldsymbol k$, parity is no longer a good quantum number for the finite-volume multi-particle states. We will avoid this issue by restricting attention to $\boldsymbol k = \boldsymbol 0$ in the following section.

Various strategies are possible for extracting this finite-volume estimator from the amputated four-point function. One technical issue affecting all methods is that, because 
$\hat{\Gamma}^{(4)}_\mu(t_H;\boldsymbol{p},\boldsymbol{k})_{L}$ only depends on the projected spectral functions, $\mathbb P_{p}(\boldsymbol p) \, \widetilde{\sigma}_{\mu}(\omega)_{L} \, \mathbb P_{\Sigma}(\boldsymbol k)$ and $\mathbb P_{p}(\boldsymbol p) \, \widetilde{\rho}_{\mu}(\omega)_{L} \, \mathbb P_{\Sigma}(\boldsymbol k)$, it is only possible to extract a similarly projected version of $\widetilde{F}_\mu(\boldsymbol{k},\boldsymbol{p})_{L}$. Rather than carrying these projectors in all subsequent equations, we find it most convenient to change to spin indices at this stage, defining
 \begin{align}
 F^{rs}_\mu(\boldsymbol{k},\boldsymbol{p})_{L} & = \bar{u}^r_p(\boldsymbol p) \widetilde{F}_\mu(\boldsymbol{k},\boldsymbol{p})_{L} u^s_{\Sigma}(\boldsymbol k) \,, \\
 \hat{\Gamma}^{(4)rs}_\mu(t_H;\boldsymbol{p},\boldsymbol{k})_{L}
& =
 \bar{u}^r_p(\boldsymbol p) \hat{\Gamma}^{(4)}_\mu(t_H;\boldsymbol{p},\boldsymbol{k})_{L} u^s_{\Sigma}(\boldsymbol k) \,,
 \end{align}
 and similar for all other quantities defined as Dirac matrices above.

To extract $F^{rs}_\mu(\boldsymbol{k},\boldsymbol{p})_{L}$ from $\hat{\Gamma}^{(4)rs}_\mu(t_H;\boldsymbol{p},\boldsymbol{k})_{L}$, a crucial issue that any method must address is that certain intermediate states lead to exponentially growing Euclidean time dependence. This arises because, in eq.~\eqref{eq:gammahatdef}, one is multiplying by growing exponentials depending on the energies of the incoming ${\Sigma^+}$ and outgoing proton. If the intermediate energies in the sum over $n$ are sufficiently large, then these contribute decaying exponentials that outweigh the growth. However, due to low-lying finite-volume states, $\hat{\Gamma}^{(4)rs}_\mu(t_H;\boldsymbol{p},\boldsymbol{k})_{L}$ can in principle diverge either for $t_H \to \infty$ or $t_H \to - \infty$.

To understand the point in more detail, we consider the unphysical quantity
\begin{multline}
- \int_{- \infty}^{\infty}\! \intd t_H\, e^{- \omega' \vert t_H \vert} \, \hat{\Gamma}^{(4)rs}_\mu(t_H;\boldsymbol{p},\boldsymbol{k})_{L} = \\ 
\int_0^\infty \intd \omega \, \frac{\rho^{rs}_{\mu}(\omega)_L}{ E_{\Sigma}(\boldsymbol k) - \omega' - \omega } - \int_0^\infty \intd \omega \, \frac{\sigma^{rs}_{\mu}(\omega)_L}{ \omega + \omega' - E_{p}(\boldsymbol p) } \,.
\end{multline}
Here $\omega'$ is chosen such that the integral over $t_H$ is convergent, and one finds a result that is very similar to the targeted finite-volume estimator, $F^{rs}_\mu(\boldsymbol{k},\boldsymbol{p})_{L}$. In fact, the right-hand side gives this desired quantity in the $\omega' \to 0$ limit, but this is not useful as the integral on the left-hand side is divergent if evaluated at $\omega'=0$. Physically, this expression corresponds to allowing the weak Hamiltonian to carry away energy from the system such that no on-shell intermediate states occur. It thus solves the problem of growing exponentials, but at the unacceptable cost of giving an unphysical quantity.

Two closely related options are available to reach the desired expression at $\omega' = 0$. The original proposal, introduced in refs.~\cite{Christ:2015pwa, Christ:2015aha}, is to integrate $t_H$ over a finite range of times, i.e.~over the range $t_H \in [-T_a, T_b]$. One can then remove growing exponentials as a function of $T_a$ and $T_b$ in order to extract the desired finite-volume quantity.
A closely related alternative, described in ref.~\cite{Briceno:2019opb}, is to explicitly remove the exponentials as a function of $t_H$ before integrating and then to re-introduce the missing poles in a second step. 

Here we focus on the method of refs.~\cite{Christ:2015pwa, Christ:2015aha}, defining
\begin{equation}
I^{rs}_\mu(T_a,T_b;\boldsymbol{p},\boldsymbol{k})
= (-i) \int_{-T_a}^{T_b}\! \intd t_H\,\, \hat{\Gamma}^{(4)rs}_\mu(t_H;\boldsymbol{p},\boldsymbol{k}) \,,
\end{equation}
with $-T_a < 0 < T_b$. The summed correlator has a spectral representation given by
\begin{equation}
I^{rs}_\mu(T_a,T_b;\boldsymbol{p},\boldsymbol{k}) = \int_0^\infty\intd \omega \bigg [ i \rho^{rs}_{\mu}(\omega)_L \, \frac{1 - e^{ - ( \omega - E_{\Sigma}(\boldsymbol k)) T_a } }{E_{\Sigma}(\boldsymbol k)-\omega}
- i
\sigma^{rs}_{\mu}(\omega)_L \, \frac{1 - e^{ - ( \omega - E_p(\boldsymbol p) ) T_b } }{ \omega - E_p(\boldsymbol p) } \bigg ]
\,,
\label{eq:specrepintfour}
\end{equation}
in which the growing exponentials are displayed explicitly. From these expressions one sees that contributions growing as $T_a \to \infty$ will arise if $\rho^{rs}_{\mu}(\omega)_L$ includes finite-volume energies for which $E_n(\boldsymbol k) < E_{\Sigma}(\boldsymbol k)$ and similarly contributions growing as $T_b \to \infty$ will arise if $\sigma^{rs}_{\mu}(\omega)_L$ includes finite-volume energies for which $E_n(\boldsymbol p) < E_{p}(\boldsymbol p)$.

By studying the contributions in this specific system, we deduce that the limit $T_{b}\rightarrow\infty$ can be taken without any difficulties, since all possible baryonic intermediate states with strangeness $S=-1$ and momentum $\boldsymbol{p}$ have energies $E_n(\boldsymbol p)>E_p(\boldsymbol{p})$. This will be true so long as the strange quark mass is greater than the down quark mass. As a result, the term with $e^{-T_b(\omega-E_p(\boldsymbol{p}))}$ is exponentially suppressed for large $T_b$.
However, any intermediate state with $S=0$ that has an energy smaller than the initial state energy $E_{\Sigma}(\boldsymbol{k})$ will lead to an exponentially growing term when $T_a\rightarrow \infty$. For simulations at physical or close-to-physical quark masses, such intermediate states can be either a single proton state with momentum $\boldsymbol{k}$ or a nucleon-pion state with total momentum $\boldsymbol{k}$ and an energy smaller than $E_{\Sigma}(\boldsymbol{k})$. For sufficiently large light-quark masses, the energy of the finite-volume nucleon-pion states will be greater than the $\Sigma^+$ energy and become decaying exponentials in $T_a$, leaving only the single proton state to grow as $T_a \to \infty$.

To extract $F^{rs}_\mu(\boldsymbol k, \boldsymbol p)_L$ from $I^{rs}_\mu(T_a,T_b;\boldsymbol{p},\boldsymbol{k})$, all growing terms need to be removed. It is additionally possible to remove decaying states such that the $T_a \to \infty$ limit is saturated for smaller values of $T_a$. The removal of slowly decaying states was already applied in ref.~\cite{Christ:2016mmq}, in the context of rare kaon decays.
To express this compactly, it is convenient to introduce a modification of $\rho^{rs}_\mu$ that is cut off to only include low-lying states
\begin{equation}
\rho^{ [N]rs}_{\mu}(\omega)_L = \sum_{n=0}^{N-1} \delta(\omega - E_n(\boldsymbol k)) \, \frac{ C^{rs}_{n, \mu}(\boldsymbol k) }{2 E_{n}(\boldsymbol k) } \,,
\end{equation}
where $C_{n,\mu}(\boldsymbol k) $ is defined in eq.~\eqref{eq:Cdef} above, and $N$ must satisfy the condition that $E_n(\boldsymbol k) > E_{\Sigma}(\boldsymbol k) $ for $n \geq N$. We stress here that $n=0$ refers to the finite-volume single-proton state. Then one can write
\begin{align}
\label{eq:subtractedcorr}
\overline{I}^{rs}_\mu(T_a,T_b;\boldsymbol{p},\boldsymbol{k}) & = I^{rs}_\mu(T_a,T_b;\, \boldsymbol{p},\boldsymbol{k}) - \Delta{I}^{rs}_\mu(T_a;\boldsymbol{p},\boldsymbol{k}) \,, \\[5pt]
\Delta{I}^{rs}_\mu(T_a;\boldsymbol{p},\boldsymbol{k}) & \equiv (- i) \int_0^\infty\intd \omega \, \rho^{ [N]rs}_{\mu}(\omega)_L \, \frac{ e^{ - ( \omega - E_{\Sigma}(\boldsymbol k)) T_a } }{E_{\Sigma}(\boldsymbol k)-\omega} \,.
\label{eq:deltaI_def}
\end{align}
Note that  $\overline{I}^{rs}_\mu(T_a,T_b;\boldsymbol{p},\boldsymbol{k}) $ then has the desired large $T_{a,b}$ limits
\begin{equation}
\label{eq:TabF}
F^{rs}_\mu(\boldsymbol{k},\boldsymbol{p})_{L} = \,\lim_{T_{a,b}\rightarrow\infty}\overline{I}^{rs}_\mu(T_a,T_b;\boldsymbol{p},\boldsymbol{k}) \,.
\end{equation}

In contrast to $I^{rs}_\mu$, the separate quantities $\overline{I}^{rs}_\mu$ and $\Delta I^{rs}_\mu$, as well as $F^{rs}_\mu(\boldsymbol{k},\boldsymbol{p})_{L}$, have poles as a function of $L$ for any fixed kinematics. The distinction arises because, in the original expression, $1 - e^{- ( E_n(\boldsymbol k) - E_{\Sigma}(\boldsymbol k)) T_a}$ vanishes whenever $E_n$ and $E_{\Sigma}$ coincide so that the combination has a finite limit
\begin{equation}
\lim_{E_n \to E_{\Sigma}} \frac{1 - e^{ - ( E_n(\boldsymbol k) - E_{\Sigma}(\boldsymbol k)) T_a } }{E_{\Sigma}(\boldsymbol k)- E_n(\boldsymbol k)} = T_a \,.
\end{equation}
This must be the case since the manifestly finite correlator $\hat \Gamma^{(4)rs}_\mu$, integrated over a finite range of times, cannot diverge for any $L$. By contrast, $\Delta I^{rs}_\mu$, $\overline{I}^{rs}_\mu$ and $F^{rs}_\mu$ are divergent if $E_{\Sigma}(\boldsymbol k)$ coincides with a finite-volume energy. We stress that there is no problem here, this is simply part of the correct definition of the finite-volume estimator. These poles will be removed in the final relation between $F^{rs}_\mu$ and the infinite-volume Minkowski amplitude ${\mathcal A}^{rs}_\mu$.

A slight variation in extracting $ F^{rs}_\mu(\boldsymbol{k},\boldsymbol{p})_{L}$, discussed in ref.~\cite{Briceno:2019opb}, is to remove the growing exponentials before integrating. 
In the present context it leads one to define
\begin{multline}
\label{eq:IgreaterthanN}
I^{\geq N,rs}_\mu(T_a,T_b;\boldsymbol{p},\boldsymbol{k})
\equiv (-i) \int_{-T_a}^{T_b}\! \intd t_H \bigg [ \hat{\Gamma}^{(4)rs}_\mu(t_H;\boldsymbol{p},\boldsymbol{k}) \\
 - \Theta(-t_H) \int_0^\infty\intd \omega \, \rho^{[N]rs}_{\mu}(\omega)_{L} \, e^{-t_H\left[ E_{\Sigma}(\boldsymbol k) - \omega \right]} \bigg ] \,.
\end{multline}
This object now has a well-defined $T_{a,b} \to \infty$ limit, but it is not the desired estimator as the poles from the subtracted states are completely absent. These are then re-introduced by the relation
\begin{equation}
\label{eq:alternativeF}
F^{rs}_\mu(\boldsymbol{k},\boldsymbol{p})_{L} = \lim_{T_{a,b}\rightarrow\infty} I^{\geq N,rs}_\mu(T_a,T_b;\boldsymbol{p},\boldsymbol{k}) + F^{[N]rs}_\mu(\boldsymbol{k},\boldsymbol{p})_{L} \,,
\end{equation}
where we have introduced
\begin{equation}
F^{[N]rs}_\mu(\boldsymbol{k},\boldsymbol{p})_{L}
\equiv i
\int_0^\infty \intd \omega \, \frac{\rho^{[N]rs}_{\mu}(\omega)_L}{ E_{\Sigma}(\boldsymbol k) -\omega } = i \sum_{n=0}^{N-1} \, \frac{C^{rs}_{n, \mu}(\boldsymbol k)}{2 E_{n}(\boldsymbol k) \big ( E_{\Sigma}(\boldsymbol k) - E_{n}(\boldsymbol k) \big ) } \,.
\end{equation}
This method is analogous to the approach of using low-lying states to estimate the $T \to \infty$ integral for the hadronic-vacuum-polarization contribution to the magnetic moment of the muon \cite{\gmtwopapers}. 

Whether the removal of growing (and slowly decaying) exponentials is performed before or after $t_H$ integration, it requires determination of the overlaps $C^{rs}_{n, \mu}(\boldsymbol k)$ and energies $E_n(\boldsymbol k)$ of all states to be removed. We discuss the detailed approach for determining this information
in the following subsections. Having completed this, it is equally important to understand how to relate $F^{rs}_\mu(\boldsymbol{k},\boldsymbol{p})_{L}$ to the physical amplitude $\mathcal{A}^{rs}_\mu(k, p)$. This requires treating the multi-hadron finite-volume effects and understanding how to include the $N \pi$ branch cut that is part of the physical amplitude's definition. The method is discussed in detail in section~\ref{sec:FV}. In addition, the general method for extracting the form factors from the physical amplitude $\mathcal{A}^{rs}_\mu(k,p)$ is given in Appendix \ref{sec:traces}, as well as an example for a specific kinematic setup.

\subsubsection{Removal of the single-proton state}

In this subsection, we describe two methods for removing the growing exponential arising from the single-proton state.
Recalling the definitions
\begin{align}
 \big<p(\boldsymbol{p}),r\big|J_\mu(0)\big|p(\boldsymbol{k}), r' \big>_\infty & \equiv  \bar{u}^r_p(\boldsymbol{p}) \,  \widetilde{\mathcal{A}}_{\mu, p}(\boldsymbol{q})  \, u^{r'}_p(\boldsymbol{k})   \,, \\
 \big<p(\boldsymbol{k}),r' \big| \mathcal H_W(0)\big|\Sigma^+(\boldsymbol{k}),s\big>_\infty & \equiv  \bar{u}^{r'}_p(\boldsymbol{k})  \,  \widetilde{\mathcal{A}}_{H} \,  u^{s}_{\Sigma}(\boldsymbol{k})  \,,
 \end{align}
 denoting the single-proton state with a $p$ subscript (i.e.~$C^{rs}_{p,\mu}(\boldsymbol k)=C^{rs}_{n=0,\mu}(\boldsymbol k)$) one can show
\begin{align}
C^{rs}_{p,\mu}(\boldsymbol k) & \equiv \sum_{r'} \big < p(\boldsymbol{p}),r \big| J_\mu(0) \big | p(\boldsymbol k), r' \big> \, \big < p(\boldsymbol k), r' \big | {\mathcal H_{W}}(0) \big| \Sigma^+(\boldsymbol{k}),s\big> \,, \\
& = 2 M_p \, \bar{u}^{r}_p(\boldsymbol{p}) \, \widetilde{\mathcal{A}}_{\mu,p}(\boldsymbol{q}) \,  \mathbb P_p(\boldsymbol k) \, \widetilde{\mathcal{A}}_H \, u^{s}_{\Sigma}(\boldsymbol{k})\,, \\[5pt]
& =2 M_p \, \bar{u}^{r}_p(\boldsymbol{p}) \cdot \hat{\Gamma}^{(3)}_{\mu,p}(0; \boldsymbol{p},\boldsymbol{k})\cdot \hat{\Gamma}^{(3)}_H(t_H ; \boldsymbol{k}) \cdot  u^{s}_{\Sigma}(\boldsymbol{k}) e^{ t_H\left[ E_{\Sigma}(\boldsymbol k) - E_p(\boldsymbol k) \right]}  \,,
\label{eq:Cdef_pImpl}
\end{align}
where in the last line we have used the hatted three-point functions defined in eqs.~\eqref{eq:three_vec_hat} and \eqref{eq:three_hamil_hat} and have also applied the identity $\mathbb P_p(\boldsymbol p)^2 = \mathbb P_p(\boldsymbol p)$.
This expression for $C^{rs}_{p,\mu}(\boldsymbol k)$ can then be used to remove the single-particle state. One can define
\begin{align}
\Delta{I}^{(p)rs}_\mu(T_a;\boldsymbol{p},\boldsymbol{k}) & \equiv (- i)  \, \frac{ C^{rs}_{p,\mu}(\boldsymbol k)}{2 E_p(\boldsymbol k)  } \, \frac{ e^{ - ( E_{p}(\boldsymbol k) - E_{\Sigma}(\boldsymbol k)) T_a } }{E_{\Sigma}(\boldsymbol k)- E_p(\boldsymbol k)} \,,
\end{align}
as the single-proton contribution to $\Delta{I}^{rs}_\mu(T_a;\boldsymbol{p},\boldsymbol{k}) $, defined in eq.~\eqref{eq:deltaI_def}.

In fact, it is instructive here to consider the case where only the single proton leads to a growing exponential, as would be the case for sufficiently large pion mass calculations. Then the method for extracting $F^{rs}_\mu(\boldsymbol{p},\boldsymbol{k})_L$, with the proton removed after summation, can be summarized succinctly via
\begin{equation}
\label{eq:fv_one_state}
F^{rs}_\mu(\boldsymbol{p},\boldsymbol{k})_L
=   i \frac{C^{rs}_{p,\mu}(\boldsymbol k)}{2 E_p(\boldsymbol k)  } \, \frac{ e^{ T_a (E_{\Sigma}(\boldsymbol k) -  E_{p}(\boldsymbol k) )   } }{E_{\Sigma}(\boldsymbol k)- E_p(\boldsymbol k)}  -i 
  \int_{-T_a}^{\infty}\! \intd t_H  \,  \hat{\Gamma}^{(4)rs}_\mu(t_H;\boldsymbol{p},\boldsymbol{k})\,.
 \end{equation}
The estimator with the proton removed before summation instead gives 
\begin{multline}
\label{eq:other_one_state}
F^{rs}_\mu(\boldsymbol{p},\boldsymbol{k})_L
=  i  \, \frac{C^{rs}_{p,\mu}(\boldsymbol k)}{2 E_p(\boldsymbol k)  } \, \frac{ 1  }{E_{\Sigma}(\boldsymbol k)- E_p(\boldsymbol k)}  \\
-i \int_{-\infty}^{\infty}\! \intd t_H \bigg [ \hat{\Gamma}^{(4)rs}_\mu(t_H;\boldsymbol{p},\boldsymbol{k}) -  \, \Theta(- t_H)    \frac{C^{rs}_{p,\mu}(\boldsymbol k) \, e^{-t_H\left[ E_{\Sigma}(\boldsymbol k) - E_p(\boldsymbol k) \right]} }{2 E_p(\boldsymbol k)   }    \bigg ] \,,
\end{multline}
where one can directly make use of the time-dependence arising within eq.~\eqref{eq:three_hamil_hat}
\begin{equation}
C^{rs}_{p,\mu}(\boldsymbol k) \, e^{-t_H\left[ E_{\Sigma}(\boldsymbol k) - E_p(\boldsymbol k) \right]} = 2 M_p \, \bar{u}^{r}_p(\boldsymbol{p})  \cdot \hat{\Gamma}^{(3)}_{\mu,p}( 0; \boldsymbol{p},\boldsymbol{k})\cdot \hat{\Gamma}^{(3)}_H(t_H ; \boldsymbol{k}) \cdot u^{s}_{\Sigma}(\boldsymbol{k}) \,.
\end{equation}
One can readily show that these two expressions are mathematically equivalent. 
They may, however, lead to statistical differences depending on how exactly first term in eq.~\eqref{eq:fv_one_state} and the first and last terms in eq.~\eqref{eq:other_one_state} are estimated.

\bigskip

{

Note also that it follows from eqs.~\eqref{eq:fv_one_state} and \eqref{eq:other_one_state} that, in the case where $C^{rs}_{p,\mu}(\boldsymbol{k}) = 0$, there is no need to explicitly treat the single-proton state. As was shown in refs.~\cite{Christ:2015aha,Christ:2016mmq,Boyle:2022ccj}  for the single pion intermediate state in $K \to \pi \ell^+ \ell^-$, it is in fact possible to define a modified weak Hamiltonian, denoted $\mathcal H'_W(0)$, such that $F^{rs}_\mu(\boldsymbol{p},\boldsymbol{k})_L$ is invariant under $\mathcal H_W \to \mathcal H'_W$, but
\begin{align}
  \bra{p(\boldsymbol{k}),r} \mathcal{H}'_W(0) \ket{\Sigma^+(\boldsymbol{k}),s} = 0 \quad \Longrightarrow \quad C'^{\,rs}_{p,\mu}(\boldsymbol{k}) = 0 \,,
\end{align}
where the prime on $C^{rs}_{p,\mu}$ indicates the $\mathcal H_W \to \mathcal H'_W$ replacement.
The redefined Hamiltonian density is given explicitly by
\begin{equation}
  \mathcal H_W'(x) = \mathcal H_W(x) - c_S \mathcal S^{\bar d s}(x) - c_P \mathcal P^{\bar d s}(x) \,,
\end{equation}
where $c_S$ and $c_P$ are constants to be determined and
\begin{equation}
  \mathcal  S^{\bar{q}' q}(x) = \bar{q}'(x) q(x) \,, \qquad  \mathcal P^{\bar{q}' q}(x) = \bar{q}'(x) \gamma_5 q(x) \,,
\end{equation}
are flavour non-singlet scalar and pseudo-scalar densities.
In the following we first prove that $F^{rs}_\mu(\boldsymbol{p},\boldsymbol{k})_L$ is invariant under $\mathcal H_W \to \mathcal H'_W$ and then explain how one fixes $c_S$ and $c_P$ to set the single-proton contribution to vanish.

Begin by recalling that the conserved and partially conserved vector and axial currents, $V_\mu^{\bar{q}' q}$ and $A_\mu^{\bar{q}' q}$ respectively, exactly satisfy the chiral Ward identities:
\begin{equation}
\partial^\mu V_\mu^{\bar{q}' q}(x)   = i(m'-m) \mathcal{S}^{\bar{q}' q}(x) \,, \qquad
\partial^\mu A_\mu^{\bar{q}' q}(x)   = i(m'+m) \mathcal{P}^{\bar{q}' q}(x) \,.
\end{equation}
 Thus, inserting the scalar and pseudo-scalar densities between generic final and initial states with matching momenta, $ \bra{E_f, \boldsymbol k}_L$ and $ \ket{E_i, \boldsymbol k}_L$ respectively, one finds
\begin{align}
  \label{eq:scalarWI}
	  \bra{E_f, \boldsymbol k} \mathcal{S}^{\bar{q}' q}(x) \ket{E_i, \boldsymbol k}_L = i \frac{E_i-E_f}{m'-m}  \bra{E_f, \boldsymbol k}  V_0^{\bar{q}' q}(x) \ket{E_i, \boldsymbol k}_L \,, \\
	  \bra{E_f, \boldsymbol k} \mathcal{P}^{\bar{q}' q}(x)\ket{E_i, \boldsymbol k}_L = i \frac{E_i-E_f}{m'+m}  \bra{E_f, \boldsymbol k} A_0^{\bar{q}' q}(x) \ket{E_i, \boldsymbol k}_L \,.
    \label{eq:pseudoscalarWI}
\end{align}

The crucial point is that $\mathcal S^{\bar d s}(x)$ and $\mathcal P^{\bar d s}(x) $ have no effect on $F_\mu^{rs}(\boldsymbol p, \boldsymbol k)_L$. 
To demonstrate this we define $F'^{\,rs}_{\mu}(\boldsymbol p, \boldsymbol k)_L$ as the result of replacing $\mathcal H_W \to \mathcal H'_W$ in $F^{\,rs}_{\mu}$, and then use the $c_S$ and $c_P$ dependence to unambiguously decompose as
\begin{equation}
  F'^{\,rs}_{\mu}(\boldsymbol p, \boldsymbol k)_L = F_{\mu}^{rs}(\boldsymbol p, \boldsymbol k)_L - c_S S_{\mu}^{rs}(\boldsymbol p, \boldsymbol k)_L - c_P P_{\mu}^{rs}(\boldsymbol p, \boldsymbol k)_L \,,
\end{equation} 
thereby defining $ S_{\mu}^{rs}(\boldsymbol p, \boldsymbol k)_L $ and $ P_{\mu}^{rs}(\boldsymbol p, \boldsymbol k)_L$. Taking the scalar for concreteness note that this can then be written explicitly as
\begin{multline}
  S_\mu^{rs}(\vect{p},\vect{k})_L =  \ i \sum_{{n'} }  \frac{1}{2 E_{{n'}}(\boldsymbol k)} \frac{\bra{p(\vect{p}),r} J_\mu \ket{E_{n'},\vect{k}}_L \bra{E_{n'},\vect{k}} \mathcal{S}^{\bar{d}s} \ket{\Sigma^+(\vect{k}),s}_L}{E_{\Sigma}(\vect{k})-E_{{n'}}(\boldsymbol k)} \\
   -i \sum_{n}  \frac{1}{2 E_{n}(\boldsymbol p)}  \frac{\bra{p(\vect{p}),r} \mathcal{S}^{\bar{d}s} \ket{E_n,\vect{p}}_L \bra{E_n,\vect{p}} J_\mu \ket{\Sigma^+(\vect{k}),s}_L}{E_n(\vect{p})-E_p(\vect{p})} \,,
\end{multline}
where the sum over $n'$ runs over finite-volume states with proton quantum numbers and that over $n$ runs over states with strangeness $S = -1$. Using eq.~\eqref{eq:scalarWI} then gives
  \begin{align} 
    \begin{split}
    S_\mu^{rs}(\vect{p},\vect{k})_L & =  \ \frac{1}{m_s-m_d} \left( \sum_{{n'} } \frac{1}{2 E_{{n'}}(\boldsymbol k)} \bra{p(\vect{p}),r} J_\mu \ket{E_{n'},\vect{k}}_L \bra{E_{n'},\vect{k}} V^{\bar{d}s}_0 \ket{\Sigma^+(\vect{k}),s}_L \right. \\
    & \left. \hspace{5em} - \sum_{{n} }  \frac{1}{2 E_{{n}}(\boldsymbol p)} \bra{p(\vect{p}),r} V^{\bar{d}s}_0 \ket{E_{n},\vect{p}}_L \bra{E_{n},\vect{p}} J_\mu \ket{\Sigma^+(\vect{k}),s}_L \right) \,,
  \end{split} 
\end{align}
where the energy differences in the chiral Ward identity have cancelled the poles.
As a result each term now contains an insertion of the identity that can be collapsed to reach
\begin{align}
  S_\mu^{rs}(\vect{p},\vect{k})_L & =  \ \frac{\bra{p(\vect{p}),r} [J_\mu ,V^{\bar{d}s}_0] \ket{{\Sigma^+}(\vect{k}),s}}{m_s-m_d}   \,.
\end{align}

Similarly, for the pseudoscalar one finds
\begin{align}
    P_\mu^{rs}(\vect{p},\vect{k})_L = & \ \frac{\bra{p(\vect{p}),r} [J_\mu ,A^{\bar{d}s}_0] \ket{\Sigma^+(\vect{k}),s}}{m_s+m_d} \,.
\end{align}
Since the electromagnetic current $J_\mu$ is a singlet in flavour space and the flavour changing axial and vector currents are not, they will commute causing the shifts to the estimator to vanish. 
We can therefore shift the weak Hamiltonian by any amount of the scalar and pseudoscalar operators without affecting the value of the estimator:
\begin{equation}
  S_\mu^{rs}(\vect{p},\vect{k})_L = 0 \,, \quad
    P_\mu^{rs}(\vect{p},\vect{k})_L = 0   \qquad \Longrightarrow \qquad    F'^{\,rs}_{\mu}(\boldsymbol p, \boldsymbol k)_L = F_{\mu}^{rs}(\boldsymbol p, \boldsymbol k)_L \,.
\end{equation}

It remains only to fix the values of $c_S$ and $c_P$.
This can be achieved by demanding
\begin{align}
\bra{p(\boldsymbol{k}),r} \mathcal{H}^{' \pm}_W \ket{\Sigma^+(\boldsymbol{k}),s} = 0 \,,
\end{align}
where we have restored the $\pm$ superscript as it is relevant here that one can study the definite parity sectors separately.
As shown in appendix \ref{sec:appFFdec}, a generic Lorentz (pseudo)scalar operator $\mathcal{O}$ can be decomposed into scalar and pseudoscalar form factors $a_{\mathcal{O}}$ and $b_{\mathcal{O}}$ respectively, giving the decompositions
\begin{align}
\bra{p(\boldsymbol{k}),r} \mathcal{H}^{'+}_W \ket{\Sigma^+(\boldsymbol{k}),s} = & \, \bar{u}^r_p(\boldsymbol{k}) \left(a_H - c_S \, a_S \right) u^s_{\Sigma}(\boldsymbol{k}) \,, \\
\bra{p(\boldsymbol{k}),r} \mathcal{H}^{'-}_W \ket{\Sigma^+(\boldsymbol{k}),s} = & \, \bar{u}^r_p(\boldsymbol{k}) \left(b_H - c_P \, b_P \right) \gamma_5 \, u^s_{\Sigma}(\boldsymbol{k}) \,,
\end{align} 
where we have used that only $\mathcal S^{\bar d s}$ contributes to $\mathcal H_W^+$ (and only $\mathcal P^{\bar d s}$ to $\mathcal H_W^-$).
We deduce $c_S = \frac{a_S}{a_H}$ and $c_P = \frac{b_P}{b_H}$ are required for the single proton intermediate state to vanish. 

Comparing this to the scalar shift in the rare kaon decay \cite{Christ:2015aha}, we note that due to the additional spin degree of freedom, there is an additional pseudoscalar form factor describing the weak Hamiltonian matrix element. In general, it is therefore not sufficient to perform only a scalar shift to remove the single proton intermediate state from both parity sectors. There is however a kinematic point where the pseudoscalar shift is no longer required. This is when the spinor contraction $\bar{u}^r_p(\boldsymbol{k}) \gamma_5 u^s_{\Sigma}(\boldsymbol{k})$ vanishes at $\boldsymbol{k} = \boldsymbol{0}$, corresponding to the $\Sigma^+$ at rest.


}

\subsubsection{Removal of multi-hadron states}
At close-to-physical quark masses, the other growing exponentials come from the lowest-lying $N \pi$ states, which have energy smaller than the mass of the $\Sigma^+$. Because these are excited states, their subtraction is more involved. In general, all lattice interpolating operators with the correct quantum numbers will overlap the states of interest, but in practice one can only reliably extract excited state energies and matrix elements by solving a generalized eigenvalue problem (GEVP) with a diverse set of multi-hadron operators. In the present case, the low-lying states are expected to be $N \pi$ like states together with resonances. Therefore, operators built from a nucleon and pion that are individually momentum projected are required to reliably determine the excited spectrum. Depending on the finite-volume box size, three-particle states could also become important, requiring an even more complicated operator basis. 

It is also important to break up the intermediate excited states according to irreducible representations (irreps) of the octahedral symmetry group (including parity) or, in the case that the $\Sigma^+$ baryon has non-zero spatial momentum $\boldsymbol k$, to a little group that leaves the latter invariant. The GEVP analysis is then performed separately within each irrep and the final result is constructed from the separately determined spectra. Details are given in Appendix \ref{app:details_fvstates}. 

Once an optimized operator for a given excited state is determined, the removal of that state proceeds as in eq.~\eqref{eq:Cdef} for the single proton. The energy $E_n(\boldsymbol k)$ is extracted from the two-point function and the product of matrix elements
\begin{align}
C^{rs}_{n,\mu}(\boldsymbol k) & \equiv \sum_{r'} \,     \big < p(\boldsymbol{p}),r \big| J_\mu(0) \big | E_{n}, \boldsymbol k , r' \big>_L \, \big < E_{n}, \boldsymbol k, r' \big | {\mathcal H_{W}}(0) \big|\Sigma^+(\boldsymbol{k}),s\big>_L  \,,
\end{align}
from the three-point functions. It should be noted that this procedure can be applied to remove the exponentials associated with arbitrarily many intermediate states (whether growing or decaying), and for states with higher numbers of hadrons, so long as a sufficient operator basis can be obtained.

\bigskip

This completes our discussion of the construction of $F^{rs}_\mu(\boldsymbol k, \boldsymbol p)_L$. We now turn to the formalism required to relate this object to the physical amplitude for the weak decay $\sigmatopll$.


\section{Finite-volume effects}
\label{sec:FV}

For calculations with sufficiently heavy pions, the energy of the lowest $N \pi$ states will lie above the $E_\Sigma(\mathbf{k})$ threshold, and therefore only exponentially suppressed finite volume effects will be present. At the physical point however, the low-lying $N \pi$ states will be below this threshold, inducing additional power-like finite volume effects.

In this section, we detail the correction of these power-like finite volume effects from such $N \pi$ states which can be accounted for via a simple additive term, denoted by $\Delta F^{rs}_{\mu}(\boldsymbol{k},\boldsymbol{p})_L $. This allows one to determine the physical amplitude, up to exponentially suppressed $L$ effects, using the relation
\begin{align}
\label{eq:AmplitudeFVSplit}
\mathcal{A}^{rs}_\mu(k,p) & = F^{rs}_\mu(\boldsymbol{k},\boldsymbol{p})_{L} + \Delta F^{rs}_{\mu}(\boldsymbol{k},\boldsymbol{p})_L \,.
\end{align}

At this stage, we also switch from labelling states by definite individual quark flavour content (e.g.~$p$ and $\Sigma^+$) to isospin state labels ($N$ for the neutron-proton doublet and $N \pi$ for two-particle states with possible isospin values $I=1/2, 3/2$). This is necessary to introduce the finite-volume formalism for multi-particle states below.

Since $\mathcal{A}^{rs}_\mu(k,p)$ does not contain poles associated with the finite-volume $N \pi$ intermediate states, these must cancel between the two terms on the right-hand side.
As discussed in detail in ref.~\cite{Briceno:2019opb}, this means that the numerical steps that introduce poles in $F^{rs}_\mu(\boldsymbol{k},\boldsymbol{p})_{L}$ must match those in the construction of $\Delta F^{rs}_{\mu}(\boldsymbol{k},\boldsymbol{p})_L$, such that the singularities exactly cancel. To keep track of this, it is useful to group the cancelling poles 
via the definitions
\begin{align}
\mathcal{A}^{rs}_\mu(k,p) & = \lim_{T_{a,b}\rightarrow\infty} \bigg [ I^{rs}_\mu(T_a,T_b;\, \boldsymbol{p},\boldsymbol{k}) +\delta \overline F^{rs}_\mu(T_a; \boldsymbol{k},\boldsymbol{p})_L \bigg ] \,, \label{eq:AtildeFromI} \\[5pt]
\delta \overline F^{rs}_\mu(T_a; \boldsymbol{k},\boldsymbol{p})_L & = - \Delta{I}^{rs}_\mu(T_a;\boldsymbol{p},\boldsymbol{k}) + \Delta F^{rs}_\mu(\boldsymbol{k},\boldsymbol{p})_L \,.
\label{eq:deltaFBarDef}
\end{align}
An important subtlety here is that $ I^{rs}_\mu(T_a,T_b;\, \boldsymbol{p},\boldsymbol{k}) $ and $\delta \overline F^{rs}_\mu(T_a; \boldsymbol{k},\boldsymbol{p})_L $ each diverge as $T_a \to \infty$, in such a way that the combination in eq.~\eqref{eq:AtildeFromI} remains finite. 

An alternative to eq.~\eqref{eq:AtildeFromI} can be written by directly using the quantity $I^{\geq N,rs}_\mu(T_a,T_b;\boldsymbol{p},\boldsymbol{k})$, defined in eq.~\eqref{eq:IgreaterthanN}. In particular, combining the various definitions given above, one finds
\begin{align}
\mathcal{A}^{rs}_\mu(k,p) & =    \lim_{T_{a,b}\rightarrow\infty}I^{\geq N,rs}_\mu(T_a,T_b;\boldsymbol{p},\boldsymbol{k})   +\delta  F^{rs}_\mu(  \boldsymbol{k},\boldsymbol{p})_L  \,,
\end{align}
where we have introduced a $T_a$-independent analogue of $\delta \overline F^{rs}_\mu( T_a; \boldsymbol{k},\boldsymbol{p})_L$, defined as
\begin{align}
\delta F^{rs}_\mu(\boldsymbol{k},\boldsymbol{p})_L = \delta \overline F^{rs}_\mu(T_a; \boldsymbol{k},\boldsymbol{p})_L - i \sum_{n=0}^{N-1} \frac{C^{rs}_{n,\mu} (\boldsymbol{k}) [e^{-(E_n(\boldsymbol{k})-E_{\Sigma}(\boldsymbol{k}))T_a} -1 ]}{2E_n(\boldsymbol{k})(E_{\Sigma}(\boldsymbol{k})-E_n(\boldsymbol{k}))}  \,,
\label{eq:deltaFdef}
\end{align}
This is the unique combination in this work that is (i) $T_a$-independent, (ii) free of finite-volume singularities, and (iii) depends only on the states explicitly removed and not on the full sum over all states in the spectral decomposition. For this reason it will be useful in our detailed discussion of the single-channel case below.

We are now ready to give the full definition for $\Delta F^{rs}_\mu(\boldsymbol{k},\boldsymbol{p})_L$. The finite-volume correction can be written as \cite{Kim:2005gf,Christ:2015aha,Briceno:2019opb}
\begin{align}
\label{eq:DeltaFDef}
\Delta F^{rs}_\mu(\boldsymbol{k},\boldsymbol{p})_L & = i \mathcal A^r_{J_\mu}(E_{\Sigma}(\boldsymbol k), \boldsymbol{k},\boldsymbol{p}) \cdot \mathcal{F}\big ( E_{\Sigma}(\boldsymbol k), \boldsymbol{k},L \big ) \cdot \mathcal A^s_{H_W}(E_{\Sigma}(\boldsymbol k), \boldsymbol k) \,,
\end{align}
where we have introduced
\begin{align}
\mathcal A^r_{J_\mu}(E_{\Sigma}(\boldsymbol k), \boldsymbol{k},\boldsymbol{p}) &=  \big<N(\boldsymbol{p}),r\big|J_\mu(0)\big|E, (N \pi)^\mathrm{in}(\boldsymbol{k} )\big> \,, \\
\mathcal A^s_{H_W}(E_{\Sigma}(\boldsymbol k), \boldsymbol k) &=   \big<E, (N \pi)^\mathrm{out}(\boldsymbol{k} )\big| \mathcal H_W(0) \big|{\Sigma}(\boldsymbol{k}),s\big>  \,, \\
\mathcal{F}(E, \boldsymbol{P},L) & = \frac{1}{F(E, \boldsymbol{P},L)^{-1}+\mathcal{M}(E_{\sf cm})} \,.
\label{eq:calFdef}
\end{align}
Here $F(E, \boldsymbol{P},L)$ is a known geometric function, reviewed in this work in the following paragraphs, and $\mathcal A^r_{J_\mu}(\boldsymbol{k},\boldsymbol{p})$, $ \mathcal A^s_{H_W}(\boldsymbol k) $, and $\mathcal M(E_{\sf cm})$ are three types of infinite-volume amplitudes, each involving $N \pi$ states.

This construction and the following discussion is similar to that given in refs.~\cite{Christ:2015pwa,Christ:2016eae}, in that case of $\pi\pi$ intermediate states in $K-\bar{K}$ mixing and the decay $K\to\pi\overline{\nu}\nu$. The formalism of that work was generalized to particles with any spin, including the present case, in ref.~\cite{Briceno:2019opb}.
The purpose of the following is to explain the relevant formalism in the specific context of $N \pi$ states, including the role of parity, spin and non-degenerate masses. The formalism for long-range matrix elements also draws on that used for scattering and transition amplitudes \cite{Luscher:1986pf,Luscher:1990ux,Lellouch:2000pv,Rummukainen:1995vs,
Kim:2005gf,Briceno:2014oea,Hansen:2012tf}.

We begin by specifying the index space used in the definition of $\Delta F^{rs}_\mu(\boldsymbol{k},\boldsymbol{p})_L$.
The product on the right-hand side of eq.~\eqref{eq:DeltaFDef} is understood as a matrix ($\mathcal F$) contracted with two vectors ($\mathcal A^r_{J_\mu}$ and $\mathcal A^s_{H_W}$) such that no hanging indices remain. The index space of this contraction is also required for the exact definitions of $F(E, \boldsymbol P, L)$ and the other quantities appearing above. The space is denoted by $J, \ell, \mu$, referring to the total angular momentum $J$, orbital angular momentum, $\ell$, total spin (in this case fixed to $1/2$), and azimuthal component of total angular momentum $\mu$. For example,
\begin{align}
 \mathcal A^r_{J_\mu}(E, \boldsymbol{k},\boldsymbol{p})_{J, \ell, \mu} & =  \big<N(\boldsymbol{p}),r\big|J_\mu(0)\big|E, (N \pi)^\mathrm{in}(\boldsymbol{k} ) , J, \ell, \mu \big> \,, \\
\mathcal A^s_{H_W}(E, \boldsymbol k)_{J, \ell, \mu} & =   \big<E, (N \pi)^\mathrm{out}(\boldsymbol{k} ), J, \ell, \mu \, \big| \mathcal H_W(0) \big|{\Sigma}(\boldsymbol{k}),s\big>   \,,
\end{align}
and the combined labels of total energy $E$, total momentum $\boldsymbol k$, and $J, \ell, \mu$ are both necessary and sufficient to exactly specify the $N \pi$ state.

We define the geometric matrix, $F(E, \boldsymbol P, L)$, for the $N \pi$ system, by first considering the quantity for two non-identical scalar particles with masses $M_\pi, M_N$ \cite{Hansen:2016qoz}. Focusing on the $\boldsymbol P = \boldsymbol 0$ case, the definition reads
\begin{align}
F_{\ell' m';\ell m}(E, \boldsymbol 0,L)= \lim_{\alpha \to 0^+} \bigg[ \frac{1}{L^3} \sum_{\boldsymbol{k}} - \int \frac{\dd^3\boldsymbol{k}}{(2\pi)^3} \bigg] \frac{4\pi Y_{\ell' m'}^*(\hat{\boldsymbol k}) Y_{\ell m}(\hat{\boldsymbol k}) e^{- \alpha(\boldsymbol k^2 - p^2)} }{2 \omega_\pi 2 \omega_N (E-\omega_\pi-\omega_N +i \epsilon)} \bigg( \frac{\vert \boldsymbol k \vert}{p} \bigg)^{\ell +\ell '}
\, ,
\end{align}
with $\hat{\boldsymbol k}=\boldsymbol{k}/|\boldsymbol{k}|$,
$\omega_\pi=\sqrt{M_\pi^2 + \boldsymbol{k}^2}$, 
$\omega_N=\sqrt{M_N^2 + \boldsymbol{k}^2}$.
We have also introduced $p$ as the magnitude of back-to-back momentum, satisfying
\begin{align}
\label{eq:pdef}
E = E_{\sf cm}=\sqrt{M_\pi^2 + {p}^2}+\sqrt{M_N^2 + {p}^2}
\, .
\end{align}
As discussed in Refs.~\cite{Luscher:1986pf,Kim:2005gf}, ultraviolet divergences in the sum-integral difference cancel so that any smooth regulator can be used in the evaluation of each. The explicit exponential included gives one option that is also convenient for numerical evaluation. As discussed in ref.~\cite{Briceno:2014oea}, a straightforward combination of Clebsch-Gordon coefficients can then be used to promote this scalar version to the final $F$ function on the full space for particles with spin:
\begin{equation}
F_{J' \ell' \mu', J \ell \mu}(E, \boldsymbol P, L) = \sum_{m,\sigma, m'} \langle \ell m;\frac 12 \sigma | J \mu \rangle \langle \ell' m';\frac 12 \sigma | J' \mu' \rangle F_{\ell'm' ;\ell m}(E, \boldsymbol P, L) \,.
\end{equation}
This is the quantity entering the definition of $\mathcal F$ in eq.~\eqref{eq:calFdef}.

The second matrix entering the $\mathcal F$ matrix is the $N \pi \to N \pi$ scattering amplitude, denoted $\mathcal M$. This can be represented on the $J \ell \mu$ index space via
\begin{align}
\mathcal{M}_{J' \ell '\mu';J \ell \mu}(E_{\sf cm}) =\delta_{J' J} \delta_{l' l}\delta_{\mu' \mu} \frac{8 \pi E_{\sf cm}}{p \cot \delta_{J,\ell}(p) - ip} \,,
\end{align}
where
\begin{equation}
E_{\sf cm}^2 = E^2 - \boldsymbol P^2 \,,
\end{equation}
defines the centre-of-mass energy and $\delta_{J,\ell}(p)$ is the scattering phase shift with quantum numbers as indicated.

This concludes our explanation of all quantities entering $\Delta F^{rs}_\mu(\boldsymbol{k},\boldsymbol{p})_L$.
The content of eq.~\eqref{eq:DeltaFDef} can be summarized as follows: By combining a determination of the $N \pi \to N\pi$ scattering amplitude with the ${\Sigma} \overset{H_W}{\longrightarrow} N \pi$ and $N \pi \overset{J_\mu}{\longrightarrow} N$ transition amplitudes, one can calculate the correction that relates the finite-volume estimator $F^{rs}_\mu(\boldsymbol{k},\boldsymbol{p})_{L}$ to the amplitude $\mathcal{A}^{rs}_\mu(k,p) $.

\subsection{Determining \texorpdfstring{$\mathcal M$}{M}, \texorpdfstring{$\mathcal A^s_{H_W}$}{AHW} and \texorpdfstring{$\mathcal A^r_{J_\mu}$}{Amu}}

So far we have made reference to poles cancelling between $F^{rs}_\mu(\boldsymbol{k},\boldsymbol{p})_{L}$ and $\Delta F^{rs}_{\mu}(\boldsymbol{k},\boldsymbol{p})_L$, but without explicitly explaining why the poles are the same between the two terms. This follows from the observation that the condition that $\mathcal F(E, \boldsymbol P, L)$ diverges is equivalent to the Lüscher quantization condition~\cite{Luscher:1986pf,Kim:2005gf}
\begin{align}
\det \big[ \mathcal{M}^{-1}(E^{\sf cm}_n) + F(E_n, \boldsymbol P,L) \big] = 0\,,
\label{eqn:quantisation}
\end{align}
where
\begin{equation}
E_n^{\sf cm} = \sqrt{E_n^2 - \boldsymbol P^2} \,.
\end{equation}
This relation allows one to constrain the amplitude $\mathcal{M}(E^{\sf cm})$ from a numerical lattice calculation by computing many finite-volume energies.

The energies, $E_n$, directly correspond to the poles in $\Delta F^{rs}_\mu(\boldsymbol k, \boldsymbol p)_L$, as can be readily seen from its definition in eq.~\eqref{eq:DeltaFDef}. In the limit of infinite statistics, the formalism guarantees a perfect cancellation between the poles in $F^{rs}_\mu(\boldsymbol k, \boldsymbol p)_L$ and those in $\Delta F^{rs}_\mu(\boldsymbol k, \boldsymbol p)_L$. However, for realistic data, with statistical uncertainties, care must be taken. In particular, given the best fit for the lattice energies $E_n(L)$ from Euclidean correlators, one can use a given parametrization in the quantization condition to determine the best fit for $\mathcal{M}(E^{\sf cm})$.  
If the lattice energies (rather than the best-fit quantization energies) are used to construct $F^{rs}_\mu(\boldsymbol k, \boldsymbol p)_L$ singularities will fail to cancel in the amplitude. As discussed in ref.~\cite{Briceno:2019opb}, the most straightforward solution here is to directly use the best-fit quantization energies everywhere.

The Lellouch-Lüscher formalism \cite{Lellouch:2000pv} and its extensions can be used to determine the remaining amplitudes $\mathcal A^s_{H_W}$ and $\mathcal A^r_{J_\mu}$. This follows from the eigenvectors of the quantization condition matrix which is known to be rank one near the finite-volume energy
\begin{equation}
\label{eq:LLdef}
\lim_{E \to E_n(L)} \big (E - E_n(L) \big )\frac{1}{ \mathcal{M}(E^{\sf cm}_n) + F^{-1}(E_n, \boldsymbol P,L) } = \mathcal E^{(n), {\sf in}} \otimes \mathcal E^{(n), {\sf out}} \,,
\end{equation}
where the $\otimes$ indicates an outer-product in the same index space, so that $\mathcal E^{(n), {\sf in}} $ and $ \mathcal E^{(n), {\sf out}}$ each carry the $J, \ell, \mu$ indices.
Then the amplitudes are related to finite-volume matrix elements via
\begin{align}
  \big < E_{n}, \boldsymbol k \big | {\mathcal H_{W}}(0) \big|{\Sigma}(\boldsymbol{k}),s\big>_L & = \mathcal E^{(n), {\sf out}} \cdot \mathcal A^s_{H_W}(E, \boldsymbol k)  \,, \\
 \big<N(\boldsymbol{p}),r\big|J_\mu(0)\big|E_{n}, \boldsymbol k \big>_L & =  \mathcal A^r_{J_\mu}(E, \boldsymbol k,\boldsymbol p) \cdot \mathcal E^{(n), {\sf in}} \,,
\end{align}
up to an inherent phase ambiguity that will cancel between $\mathcal E^{(n), {\sf out}}$ and $\mathcal E^{(n), {\sf in}}$ in the full finite-volume correction.

\subsection{Single-channel case}

We now show that, in the limit where only a single channel contributes, that this formalism is equivalent to that of ref.~\cite{Christ:2015pwa}. We first consider the general case in which $L$ is not tuned such that $E_n(L) = M_{\Sigma}$ and then examine the finely tuned case discussed in ref.~\cite{Christ:2015pwa}. We also extend the latter by expanding to higher orders about the tuned volume point. This allows one to propagate uncertainties, given that $E_n(L) = M_{\Sigma}$ will never hold exactly in estimators of the variance.

For the purpose of this discussion, we focus attention on the case of zero momentum in the finite-volume frame ($\boldsymbol P = \boldsymbol 0$). We specifically consider the case of a single $N \pi$ channel in the $P$-wave $(\ell = 1)$. In this case the matrices $\mathcal{M}(E)$ and $ F(E,L)$ are one-dimensional. It is standard to describe these in terms of the scattering phase shift $\delta^{\ell = 1}(E)$ and the so-called pseudophase $\phi(E,L)$:
\begin{equation}
\mathcal{M}(E) = \frac{8 \pi E }{p} \frac{1}{\cot \delta(E ) -i} \, , \qquad
F(E,L) = \frac{p}{8 \pi E } \, \big [ \! \cot \phi(E,L) + i \big ]
\,.
\end{equation}

Substituting this into eq.~\eqref{eq:LLdef} gives
\begin{align}
\mathcal E^{(n), {\sf in}} \mathcal E^{(n), {\sf out}} &
= - \left [ \mathcal M^2(E) \frac{\partial}{\partial E} \left( F(E, L) + \mathcal M^{-1}(E) \right ) \right ]_{E=E_n}^{-1} \,, \\[10pt]
& \hspace{0pt} = -\frac{p}{8 \pi E} \left [ \sin^2\! \delta(E)\ e^{ 2i \delta(E)}\ \frac{\partial}{\partial E} \big ( \! \cot\phi(E, L) + \cot\delta(E) \big ) \right ]_{E=E_n}^{-1} \,, \\[10pt]
& \hspace{0pt} = \frac{p}{8 \pi E} e^{-2i \delta(E)} \left [ \frac{\partial}{\partial E} \big( \phi(E, L) + \delta(E) \big ) \right ]_{E=E_n}^{-1} \,.
\end{align}
This can be used to express the relation between finite- and infinite-volume matrix elements as
\begin{multline}
\label{eq:onedimAA}
\mathcal A^r_{J_\mu}(E_n, \boldsymbol 0,\boldsymbol p)
\mathcal A^s_{H_W}(E_n, \boldsymbol 0) 
=
  \big<N(\boldsymbol{p}),r\big|J_\mu(0)\big|E_{n}, \boldsymbol 0 \big>_L \\[10pt]
\times \frac{4 \pi }{p} e^{+2i \delta(E)} \left [ \frac{\partial \phi(E,L) }{\partial E} + \frac{\partial \delta(E) }{\partial E} \right ]_{E=E_n} \big < E_{n}, \boldsymbol 0 \big | {\mathcal H_{W}}(0) \big|{\Sigma}(\boldsymbol 0),s\big>_L  
\,,
\end{multline}
where $p(E)$ is given by eq.~\eqref{eq:pdef}.
Then, in general, one can envision parametrizing the unknown function $\mathcal A^r_{J_\mu}(E, \boldsymbol 0,\boldsymbol p) \mathcal A^s_{H_W}(E, \boldsymbol 0)$ as a function of $E$ and constraining the unknown parameters using eq.~\eqref{eq:onedimAA}. 

Taking this function as known allows us to construct $ \delta F^{rs}_\mu(\boldsymbol k, \boldsymbol p)_L$ as defined in eq.~\eqref{eq:deltaFdef}. This is a useful quantity to focus on as it only depends on the states that have been removed explicitly, together with the finite-volume correction. 
\begin{multline}
\delta F^{rs}_\mu(\boldsymbol{0},\boldsymbol{p})_L = i \sum_{n=0}^{N-1} \, \frac{ C^{rs}_{n, \mu}(\boldsymbol 0)}{2 E_{n} \big ( M_{\Sigma} - E_{n} \big ) } \\
- i \frac{p_{\Sigma}}{8 \pi M_{\Sigma} } \frac{\mathcal A^r_{J_\mu}(M_{\Sigma}, \boldsymbol 0,\boldsymbol p) \mathcal A^s_{H_W}(M_{\Sigma}, \boldsymbol 0) }{ \big [ \! \cot \phi(M_{\Sigma},L) + i \big ]^{-1} + \big [ \! \cot \delta(M_{\Sigma} ) -i \big ]^{-1} }  \,,
\label{eq:Fgi_general_finalform}
\end{multline}
where $p_{\Sigma} = p(M_{\Sigma})$.
This can alternatively be written as 
  \begin{align}
    \begin{split}
    \delta F^{rs}_\mu(\boldsymbol{0},\boldsymbol{p})_L & =  i  \frac{ C^{rs}_{p, \mu}(\boldsymbol 0)}{2 M_p \big ( M_{\Sigma} - M_p \big ) } 
    \\[3pt]
    &  + i \sum_{n=1}^{N-1}   \frac{p_n}{8 \pi E_n }   \, \frac{\mathcal A^r_{J_\mu}(E_n, \boldsymbol 0,\boldsymbol p) e^{-2i \delta(E_n)} 
    \mathcal A^s_{H_W}(E_n, \boldsymbol 0) }{  \big ( M_{\Sigma} - E_{n} \big ) \partial_E [\phi(E, L) + \delta(E)] } \bigg \vert_{E = E_n} \\[3pt]
   & - i \frac{p_{\Sigma}}{8 \pi M_{\Sigma} } \mathcal A^r_{J_\mu}(M_{\Sigma}, \boldsymbol 0,\boldsymbol p) e^{- 2 i \delta(M_{\Sigma})} \mathcal A^s_{H_W}(M_{\Sigma}, \boldsymbol 0) \Big ( \! \cot \!\big [ \delta(M_{\Sigma} )+ \phi(M_{\Sigma},L) \big ]+i \Big ) \,,
    \end{split}
    \label{eq:dFgi_final}
    \end{align}
where, with the exception of the single-proton state, we have expressed also the finite-volume matrix elements within $C_{n,\mu}(\boldsymbol 0)$ in terms of the infinite-volume transition amplitudes. Here we have also introduced the shorthand $\partial_E = \partial/\partial E$ to express the derivative factor in the first term more compactly.
The result summarized in eq.~\eqref{eq:dFgi_final} matches eq.~(35) of ref.~\cite{Christ:2015pwa} up to some superficial differences in notation, motivated by the fact that we are describing a different system. For example in contrast to the case of the $K_L$-$K_S$ mass difference described in ref.~\cite{Christ:2015pwa}, here the two infinite-volume amplitudes entering the correction, $\mathcal A^r_{J_\mu}(E_n, \boldsymbol 0,\boldsymbol p) $ and $
\mathcal A^s_{H_W}(E_n, \boldsymbol 0)$, are different. For this reason we can not express the result as a magnitude-squared but instead explicitly include the Watson phase $e^{-2i \delta} $ such that the combination $\mathcal A^r_{J_\mu} e^{-2i \delta}  \mathcal A^s_{H_W}$ is real-valued (but not necessarily positive).

\bigskip

This concludes our discussion of the general finite-volume formalism. Before moving to the next section, which describes the case of a volume tuned such that $M_\Sigma$ coincides with a finite-volume energy, we close here by summarizing the necessary steps to obtain the full amplitude $\mathcal{A}^{rs}_\mu(k,p)$ with all power-like finite-volume corrections removed. 

First, one needs to compute the finite-volume estimator $\mathbb P_B(\boldsymbol p) \, \widetilde{F}_\mu(\boldsymbol{k},\boldsymbol{p})_{L}\, \mathbb P_B(\boldsymbol k)$ on the lattice, giving access to its spin-projected counterpart, $F^{rs}_\mu(\boldsymbol{k},\boldsymbol{p})_{L} $, as described in eqs.~\eqref{eq:TabF} and \eqref{eq:alternativeF}. The next step is to perform an $N \pi$ scattering analysis as has been done in refs.~\cite{Andersen:2017una,Andersen:2019ktw} and as we outline in appendix~\ref{app:details_fvstates}. This involves obtaining the finite-volume energies, which via the generalized L\"uscher formalism lead to parametrizations of the phase shift $\delta(E)$ as well as its derivative. Similarly, a lattice computation of the three-point functions on the right-hand side of eq.~\eqref{eq:onedimAA}, together with the generalized Lellouch-L\"uscher relation, gives access to the infinite-volume transition amplitudes $\mathcal A^r_{J_\mu}(E_n, \boldsymbol 0,\boldsymbol p)$ and $ \mathcal A^s_{H_W}(E_n, \boldsymbol 0) $. Putting everything together, these ingredients can now be put into eq.~\eqref{eq:dFgi_final} to obtain $\delta F^{rs}_\mu(\boldsymbol{0},\boldsymbol{p})_L$ and the final amplitude via eqs.~\eqref{eq:AtildeFromI}, \eqref{eq:deltaFBarDef}, and \eqref{eq:deltaFdef}. We stress that the final $+i$ contribution in eq.~\eqref{eq:dFgi_final} allows direct access to the imaginary part of the amplitude $\mathcal{A}^{rs}_\mu(k,p)$.

\subsection{Expanding about the pole}

The result of the previous section gives the finite-volume correction for any value of $L$ for the case of $E = M_{\Sigma}$. In this section we denote by $\bar L$ and $\bar n$ particular values such that $E_{\bar n}(\bar L) = M_{\Sigma}$ for one of the finite-volume states. The assumption that one is exactly tuned to such a volume is built into ref.~\cite{Christ:2015aha} and performing the expansion here allows us to make contact to the earlier work.

\bigskip

\textbf{Cancellation of leading-order singularity}

Defining $\delta L = L - \bar L$ we first note that the expression for $ \delta F^{rs}_\mu(\boldsymbol{0},\boldsymbol{p})_L$ has a pole that must cancel. The leading order expansion explicitly gives
\begin{multline}
\delta F^{rs}_\mu(\boldsymbol{0},\boldsymbol{p})_L = i \frac{p_{\Sigma}}{8 \pi M_{\Sigma} } \frac{\mathcal A^r_{J_\mu}(M_{\Sigma}, \boldsymbol 0,\boldsymbol p)
\mathcal A^s_{H_W}(M_{\Sigma}, \boldsymbol 0) }{\delta L }
\\[5pt]
\times \bigg [ \! - \frac{e^{-2i \delta(M_{\Sigma})}}{ \partial_{L} E_{\bar n}(L) } \left [ \partial_E \phi(E,L)  + \partial_E \delta(E) \right ]^{-1}_{E=M_{\Sigma}}
- \frac{ 1 }{ \partial_{ L} \big [ \! \cot \phi(M_{\Sigma}, L) + i \big ]^{-1} } \bigg ]_{L = \bar L}+ \mathcal O(\delta L^0) \,.
\end{multline}
To see that the term in square brackets vanishes, first use the identity
\begin{align}
0 & = \frac{d}{d L} \Big ( \phi(E_n(L),L) + \delta(E_n(L) ) \Big ) \,, \\
& = \partial_L E_n(L) \times \partial_E \Big ( \phi(E ,L) + \delta(E ) \Big ) + \partial_L \phi(E ,L) \,,
\end{align}
where we have introduced the shorthand $\partial_L = \partial/(\partial L)$. In the first line it is understood that the total derivative is being evaluated along the solution.
This allows us to rewrite the $L$ derivative of the energy to reach
\begin{multline}
\delta F^{rs}_\mu(\boldsymbol{0},\boldsymbol{p})_L = i \frac{p_{\Sigma}}{8 \pi M_{\Sigma} } \frac{\mathcal A^r_{J_\mu}(M_{\Sigma}, \boldsymbol 0,\boldsymbol p)
\mathcal A^s_{H_W}(M_{\Sigma}, \boldsymbol 0) }{\delta L }
\\
\times \bigg [ e^{- 2i \delta(M_\Sigma)} \left [\partial_{ L} \phi(E,  L) \right ]^{-1}_{E=M_{\Sigma}} 
- \frac{ 1 }{ \partial_{ L} \big [ \! \cot \phi(M_{\Sigma}, L) + i \big ]^{-1} } \bigg ]_{L = \bar L}+ \mathcal O(\delta L^0) \,.
\end{multline}
Finally, using
\begin{align}
\partial_L \bigg ( \big [ \! \cot \phi(E,L) + i \big ]^{-1} \bigg )
&= - \big [ \! \cot \phi(E,L) + i \big ]^{-2} \partial_L \cot \phi(E,L) \,, \\
& = - e^{- i 2 \phi(E, L) } \partial_L \phi(E,L) \,,
\end{align}
and replacing $\phi(M_{\Sigma}, \bar L)$ with $- \delta^{\ell = 1}(M_{\Sigma})$,
we see the desired cancellation to deduce that $\delta F_\mu $ remains finite as $\delta L \to 0$.

\bigskip

\textbf{Result for $L = \bar L$}

To derive the $(\delta L)^0$ term simply requires carefully expanding to the next order. Performing the expansion about $\delta L = 0$ leads to numerous terms, depending on both first and second derivatives of $E_{\bar n}(L)$ with respect to $L$.  The result can be written as
\begin{align}
  \begin{split}
  \delta F^{rs}_\mu(\boldsymbol{0},\boldsymbol{p})_L & =  i \sum_{n \neq \bar n}^{N-1} \, \frac{C^{rs}_{n, \mu}(\boldsymbol 0)}{2 E_{n} \big ( M_{\Sigma} - E_{n} \big ) }
  \\
& \hspace{0pt}  - i \frac{p_{\Sigma} \, e^{-2 i \delta(M_{\Sigma})} \, \mathcal A^r_{J_\mu}(M_{\Sigma}, \boldsymbol 0,\boldsymbol p) \mathcal A^s_{H_W}(M_{\Sigma} , \boldsymbol 0) }{8 \pi M_{\Sigma} \, \partial_E [ \phi(E, \bar L) + \delta(E) ] } \\ 
& \hspace{60pt}  \times
  \bigg [    
    \partial_E \log \big [ p(E) \mathcal A^r_{J_\mu}(E, \boldsymbol 0,\boldsymbol p) e^{- 2 i \delta(E)} \mathcal A^s_{H_W}(E , \boldsymbol 0) /E \big ]
  \\ 
  & \hspace{70pt} + i    \partial_E [ \phi (E,\bar L) + \delta(E) ]
  + \frac{1}{2} \frac{ \partial_E^2 [ \phi(E, \bar L) + \delta(E)]}{\partial_E [ \phi(E, \bar L) + \delta(E) ]} \bigg ]_{E = M_{\Sigma}} \\
  & \hspace{60pt}
   + \mathcal O(\delta L)\,,
  \end{split}
  \end{align}
  where we emphasize that $n=0$ (corresponding to the single-proton state) is included in the sum in the first term.
We have reached this result both by expanding about $\delta L$ (equivalently taking the $L \to \bar L$ limit) with $E = M_{\Sigma}$ fixed and, alternatively, by setting $L = \bar L$ exactly and taking the $E \to M_{\Sigma}$ limit. As expected the same result is recovered in both approaches.

\textbf{$O(\delta L)$ correction}

Finally, one can perform the tedious exercise to push this expansion to one higher order in $\delta L$. The result can be written as
\begin{equation}
\Big [ \delta F^{rs}_\mu(\boldsymbol{0},\boldsymbol{p})_L \Big ]^{\mathcal O(\delta L)} = -  i   \, e^{-2 i  \delta(M_{\Sigma})} ( M_{\Sigma} \delta L)  \frac{\mathcal A^r_{J_\mu}(M_{\Sigma}, \boldsymbol 0,\boldsymbol p) \mathcal A^s_{H_W}(M_{\Sigma} , \boldsymbol 0)        }{96 \pi   \mathcal D^4   (  \phi^{(0,1)})^2 }     \big (    \Xi   +   \mathcal X \big )  \,,
\end{equation}
where the square brackets and superscript on the left-hand side indicate that we are reporting only the linear order, to be added to the constant order above. Here we have introduced the lengthy expressions
\begin{align}
\Xi & =   \left(\phi^{(0,1)}\right)^3 \bigg [-2  \mathcal D  \Pi \phi^{(3,0)}+6 \phi^{(1,0)} \Pi \phi^{(2,0)}-3 \Pi \left(\phi^{(2,0)}\right)^2-6  \mathcal D^2 \Pi^{(2)} 
\nonumber
\\
&
\hspace{70pt}+12  \mathcal D^2 \left(1+2 i \delta^{(1)}\right) \Pi^{(1)}
+ 12 i  \mathcal D^2 \Pi \delta^{(2)}+24  \mathcal D^2 \Pi \left(\delta^{(1)}\right)^2-24 i  \mathcal D^2 \Pi \delta^{(1)}  
\nonumber
\\
&
\hspace{70pt}+ 4 \left( \mathcal D^2-3 \right)  \mathcal D^2 \Pi-2 \delta^{(3)}  \mathcal D  \Pi + 6 \phi^{(1,0)} \Pi \delta^{(2)}+3 \Pi \left(\delta^{(2)}\right)^2 \bigg ]
\nonumber
\\[5pt]
&
+  \left(\phi^{(0,1)}\right)^2  \bigg [ 6  \mathcal D  \left(  \mathcal D  \Pi \phi^{(2,1)}-2 \phi^{(1,1)} \Big [ \mathcal D  \Pi^{(1)}+\Pi \left(\delta^{(2)}-2 i  \mathcal D  \delta^{(1)} - \mathcal D +\phi^{(1,0)}\right) \Big ] \right) 
\nonumber
\\
&
\hspace{60pt} +  6 \mathcal D^2 
 \Big ( \phi ^{(0,1)} \big [ \Pi \left(4 i \mathcal A^{(1)} \delta^{(1)} + 2 \mathcal A^{(1)}-\mathcal A^{(2)}\right) -2 \mathcal A^{(1)}  \Pi^{(1)} \big]  -2 \mathcal A^{(1)} \Pi \phi^{(1,1)}  \Big )  \bigg ]
 \nonumber
 \\[5pt]
 &
 +6  \mathcal D^2 \Pi \phi^{(0,1)} \phi^{(0,2)} \left(-\phi^{(2,0)}+2 i (\mathcal D)^2+\phi^{(1,0)}\right)   -2 \mathcal D^4 \Pi \phi^{(0,3)} \,,
\end{align}
together with the shorthand 
\begin{gather}
\Pi^{(n)} = \frac{M^n_{\Sigma} \partial_E^n p_{\Sigma}(E)}{M_{\Sigma}} \bigg \vert_{E = M_{\Sigma}}  \,, \qquad \mathcal D = M_{\Sigma} \partial_E [ \phi(E, \bar L) + \delta(E) ] \bigg \vert_{E = M_{\Sigma}} \,, 
\\[5pt]
  \phi^{(n,m)} = M_{\Sigma}^{n-m} \partial^n_E \partial^m_L \phi(E, L) \bigg \vert_{E = M_{\Sigma}, \ L=\bar L}\,, \qquad \delta^{(n)} = M_{\Sigma}^n \partial_E^n \delta(E) \bigg \vert_{E = M_{\Sigma}} \,, \\[5pt]
 \mathcal A^{(n)} = M_\Sigma^n  \Big [ \mathcal A^r_{J_\mu}(M_{\Sigma}, \boldsymbol 0,\boldsymbol p) \mathcal A^s_{H_W}(M_{\Sigma} , \boldsymbol 0) \Big ]^{-1}    \frac{\partial^n}{\partial E^n}   \Big [  \mathcal A^r_{J_\mu}(E, \boldsymbol 0,\boldsymbol p) \mathcal A^s_{H_W}(E, \boldsymbol 0) \Big ]_{E = M_\Sigma}   \,.   
\end{gather}
A Mathematica version of this can be provided upon request to the authors.

Although this result looks quite cumbersome, we argue that it is in fact both conceptually straightforward and quite useful in practice. Given the value of the phase shift and its first and second derivatives at $E = M_{\Sigma}$ and $ L=\bar L$ together with the amplitudes $\mathcal A^r_{J_\mu}(M_{\Sigma}, \boldsymbol 0,\boldsymbol p)$ and $ \mathcal A^s_{H_W}(M_{\Sigma} , \boldsymbol 0) $, all information is available to express the above result as a single numerical coefficient multiplying $(M_{\Sigma} \delta L)$. If one is in the regime where the volume is sufficiently well tuned so that $\delta L^2$ terms can be neglected, then this gives the relevant correction from the finite-volume formalism due to inevitable slight mistuning. In particular, in a jackknife or bootstrap error estimate, the above expression gives a straightforward way to evaluate $\delta F^{rs}_\mu(\boldsymbol{0},\boldsymbol{p})_L $ on each sample and thus to propagate uncertainties into the physical amplitude.

%
%
\section{Summary}
\label{sec:summary}
In this paper, we described the different steps necessary to extract the $\sigmatopll$
rare decay amplitude from Euclidean lattice correlation functions. The main challenge is the radically
different contributions from on-shell intermediate states between Minkowski and Euclidean finite-volume
correlation functions. In Euclidean space-time, these states create spurious contributions in the
time-integrated correlation function that grow exponentially with the time range. In this paper we explain in detail how these contributions can be reconstructed and subtracted, which generalizes for the hyperon decays what was previously discussed in the kaon sector~\citep{Christ:2015aha,Christ:2016eae}, and more generally in~\citep{Briceno:2019opb}. Even after subtracting growing exponentials, the finite-volume estimator for the amplitude still suffers from finite-size effects following a power-law in the volume extent $L$. We presented the asymptotic volume behaviour of the finite-volume estimator
in different kinematics, and compared our result to previous approaches. We also expand on previous work by providing the leading order correction of the finite-volume effects for lattice volumes tuned to have a finite-volume energy level near the ${\Sigma}$ mass.

This work paves the way for predicting this amplitude in future lattice calculations, and we are in parallel conducting exploratory numerical simulations to that effect \cite{Hodgson:2021lzt}. Beyond the theoretical problems presented in this paper, one can expect important numerical challenges to be addressed, in particular related to the generally poor statistical signal of baryonic correlation functions in lattice calculation. However, such lattice calculation will generate numerous interesting results beyond the exciting perspective of studying this flavour-changing neutral current decay observed at LHCb. Indeed, tackling the growing exponential issue will generate as a side-product first-principle determinations of $N \pi$ scattering parameters. More generally, this type of process is a perfect ground to extend our general understanding on how to predict non-local hadronic matrix elements from lattice simulations.
%
%
%
%

\appendix
%
%
%
%
%


\section{Form-factor decomposition (Minkowski)}
\label{sec:appFFdec}

Generally the matrix element of an operator $\mathcal{O}$ between two baryonic states $\ket{B(\vect{p}),s}$ and $\ket{B'(\vect{p}'),r}$, with momenta $\vect{p}$ and $\vect{p}'$ and spin projections $s$ and $r$ can be split into the external spinors $u$ and a $4 \times 4$ Dirac-matrix amplitude $\widetilde{\mathcal{A}}$
\begin{align}
\mathcal{A}^{rs}(p,p') = \bra{B'(\vect{p}'),r} \mathcal{O} \ket{B(\vect{p}),s} = \bar{u}_{B'}^{r}(\vect{p}') \widetilde{\mathcal{A}}(q,k) u^{s}_B(\vect{p}) \,,
\end{align}
where we now parametrize in terms of the 4-momenta $q=p-p'$ and $k=p+p'$. We then write $\widetilde{\mathcal{A}}$ in terms of all of the available Dirac-matrix structures $\Gamma \in \left\{ 1, \gamma_5, \gamma_{\mu}, \gamma_{\mu}\gamma_5, \sigma_{\mu \nu} \right\}$ in such a way as the Lorentz structure of the matrix element is recovered. These terms are associated with a form factor which is a scalar coefficient, and can only be a function of the Lorentz and spin scalar objects $q^2$ and $k^2$. However, $k^2 = 2m^2 + 2m'^2 - q^2$ and therefore $k^2$ and $q^2$ are not independent, so the forms factors can be written as only a function of $q^2$.

The base set of Dirac-matrix Lorentz scalar objects is:
\begin{align}
\{ 1, \gamma_5, \slashed{q}, \slashed{q} \gamma_5, \slashed{k}, \slashed{k} \gamma_5, \sigma_{\mu \nu}q^{\mu} k^{\nu}, \sigma_{\mu \nu}q^{\mu} k^{\nu} \gamma_5 \} \,.
\end{align}
The full list of structures is the infinite set of all combinations of these base elements. These additional structures can always be decomposed into a linear combination of the base elements using $\slashed{p}\slashed{p} = p^2$ and $\slashed{p}\slashed{p}' = 2 p \cdot p' - \slashed{p}' \slashed{p}$ for any $p,p'$.

The base set of spin-matrix Lorentz vectors objects is:
\begin{align}
\{ q_{\mu}, q_{\mu} \gamma_5, k_{\mu} , k_{\mu} \gamma_5, \gamma_{\mu}, \gamma_{\mu} \gamma_5, \sigma_{\mu \nu} q^{\nu}, \sigma_{\mu \nu} q^{\nu} \gamma_5, \sigma_{\mu \nu} k^{\nu}, \sigma_{\mu \nu} k^{\nu} \gamma_5 \} \,,
\end{align}
where again the infinite combinations can be decomposed in terms of the base set.

\subsection{Explicit form-factor decompositions}
Using the base set for a generic Lorentz (pseudo-)scalar operator $\mathcal{S}$ as given above yields the form factor decomposition
\begin{align}
\mathcal{A}^{rs}(p,p') = & \bra{B'(\vect{p}'),r} \mathcal{S} \ket{B(\vect{p}), s} \\
= & \, \bar{u}^{r}_{B'}(\vect{p}') \left[ a(q^2) + b(q^2) \gamma_5 \right] u^{s}_B(\vect{p})
\, ,
\end{align}
which is relevant for the matrix element of the weak Hamiltonian in the rare hyperon decay. Analogously, using the base set for a generic Lorentz (axial-)vector operator $\mathcal{J}_{\mu}$ one obtains
\begin{align}
\mathcal{A}_{\mu}^{rs}(p,p') = & \bra{B'(\vect{p}'),r} \mathcal{J}_{\mu} \ket{B(\vect{p}),s} \\
= & \, \bar{u}^{r}_{B'}(\vect{p}') \left[ f_1(q^2) \gamma_{\mu} + i f_2(q^2) \sigma_{\nu \mu} q^{\nu} + f_3(q^2) q_{\mu} \right. \\
& \left. + g_1(q^2) \gamma_{\mu} \gamma_5 + i g_2(q^2) \sigma_{\nu \mu} q^{\nu} \gamma_5 + g_3(q^2) q_{\mu} \gamma_5 \right] u^{s}_B(\vect{p}) \,, \nonumber
\end{align}
where we have used the generalisations of the Gordon decomposition identity with initial and final states of different mass
\begin{align}
\bar{u}' \sigma_{\mu \nu} k^{\nu} u = & i (m-m') \bar{u}' \gamma_{\mu} u - i q _{\mu} \bar{u}' u \\
\bar{u}' \sigma_{\mu \nu} q^{\nu} u = & i (m+m') \bar{u}' \gamma_{\mu} u - i k_{\mu} \bar{u}' u \\
\bar{u}' \sigma_{\mu \nu} k^{\nu} \gamma_5 u = & -i (m+m') \bar{u}' \gamma_{\mu} \gamma_5 u - i q_{\mu} \bar{u}' \gamma_5 u \\
\bar{u}' \sigma_{\mu \nu} \gamma_5 q^{\nu} u = & -i (m-m') \bar{u}' \gamma_{\mu} \gamma_5 u - i k_{\mu} \bar{u}' \gamma_5 u \, .
\end{align}
Finally, the relevant amplitude for the rare hyperon decay $\Sigma^+ \to p \ell^+ \ell^-$ decomposes like
\begin{align}
& \mathcal{A}_\mu^{rs}(p,p') = \bra{p(\vect{p}'),r} \mathcal{H}_W J_{\mu} \ket{\Sigma^+(\vect{p}),s} \\
& \hspace{1cm} = \bar{u}_p^{r}(\vect{p}')  \left[ i \sigma_{\nu \mu} q^{\nu} \, ( a(q^2) + b(q^2) \gamma_5) + (q^2 \gamma_{\mu}-\slashed{q} q_{\mu}) (c(q^2) + d(q^2) \gamma_5) \right] u_{\Sigma}^{s}(\vect{p})
\end{align}
where we have used the Ward-Takahashi identity $q^\mu \, \mathcal{A}_\mu^{rs} = 0$.
Note the different definitions of momenta in this appendix and in the main text, where the $p$ carries momentum $\vect{p}$ and the $\Sigma^+$ carries momentum $\vect{k}$.


%
%
%
%
%
\section{Minkowski and Euclidean definitions and conventions}
\label{sec:appMinEuc}

In the following we give details on our conventions in Minkowski and Euclidean metric. Quantities with a sub- or super-script ``$M$'' and ``$E$'' denote quantities in Minkowski and Euclidean metric respectively\footnote{Whether the $M$ or $E$ is written as sub- or a super-script depends on notational convenience for a given quantity and does not have any meaning.}.

\subsection{Kinematic variables}
For the Minkowski metric we use the mostly-minus convention, i.e.~$g_{\mu\nu}=g^{\mu\nu}=\text{diag}(1,-1,-1,-1)$. Then the four-momentum of a state with energy $E$ and spatial momentum $\boldsymbol p$ is given by
\begin{equation}
 p_{M}^\mu = (E,\boldsymbol{p}) \,, \ \  p^M_\mu = (E,-\boldsymbol{p})\,, \qquad \text{and}\qquad p_E^\mu = p^E_\mu = (iE,\boldsymbol{p}) \,,
 \label{eq:kinfourvec}
\end{equation}
leading to
 \begin{equation}
  p_M^2 = p^M_\mu p_M^\mu = E^2-\boldsymbol{p}^2 \,, \qquad \text{and} \qquad p_E^2 = p^E_\mu p_E^\mu = -E^2 + \boldsymbol{p}^2 = -p_M^2 \,.
 \end{equation}
Of course, the limitation of the Euclidean signature in numerical lattice calculations is that one can often only access Euclidean momenta with real components, so that some effort is needed to access extract observables for which $p_E^2 < 0$.

\subsection{Gamma matrices}

We define the Euclidean $\gamma$-matrices from their Minkowski-counterparts as follows:
\begin{equation}
 \gamma^E_0 \equiv \gamma^M_0\,=\gamma_M^0 \qquad
 \gamma^E_j \equiv i\gamma^M_j\,=-i\gamma_M^j \,.
 \label{eq:euclgammamu}
\end{equation}
These $\gamma$-matrices fulfil the following anti-commutation relations
\begin{equation}
 \left\{\gamma^M_\mu,\gamma^M_\nu\right\}=2g_{\mu\nu}\mathds{1} \,, \qquad \text{and}\qquad \left\{\gamma^E_\mu,\gamma^E_\nu\right\}=2\delta_{\mu\nu}\mathds{1} \,.
\end{equation}
With these definitions $\slashed{p}$ is given by
 \begin{align}
  \slashed{p}_M & = p_M^\mu\gamma^M_\mu = p^M_\mu\gamma_M^\mu=E\gamma_M^0 - \boldsymbol{p}\cdot\boldsymbol{\gamma}^M  \,, \\
 \slashed{p}_E & = p^E_\mu\gamma^E_\mu = p^E_0\gamma^E_0 + \boldsymbol{p}\cdot\boldsymbol{\gamma}^E
 = i E\gamma_{_M}^0 - i\boldsymbol{p} \cdot \boldsymbol{\gamma}^M =i\slashed{p}_M\,.
\label{eq:euclpslash}
 \end{align}
We further define the $\gamma_5$-matrix by
\begin{equation}
 \gamma^M_5 \equiv i \gamma_M^0\gamma_M^1\gamma_M^2\gamma_M^3 = i\gamma^E_0i\gamma^E_1i\gamma^E_2i\gamma^E_3 = \gamma^E_0\gamma^E_1\gamma^E_2\gamma^E_3\equiv -\gamma_5^E\,.
\end{equation}

Finally, we introduce tensor $\sigma_{\mu\nu}$ as
\begin{equation}
 \sigma_M^{\mu\nu}\equiv\frac{i}{2}\left[\gamma_M^\mu,\gamma_M^\nu\right]=\begin{cases}
  \frac{i}{2}\left[i\gamma^E_i,i\gamma^E_j\right] = - \frac{i}{2}\left[\gamma^E_i,\gamma^E_j\right] = -\sigma^E_{ij}\hspace{1cm} \text{for}\quad \mu=i,\nu=j  \\[0.2cm] 
  \frac{i}{2}\left[i\gamma^E_i,\gamma^E_0\right] = i\cdot\frac{i}{2}\left[\gamma^E_i,\gamma^E_0\right]=i\sigma^E_{i0}\hfill\text{for}\quad \mu=i,\nu=0 \\[0.2cm]
  0 \hfill\text{for}\quad \mu=\nu
  \end{cases}
\end{equation}
where we define
\begin{equation}
 \sigma^E_{\mu\nu} \equiv \frac{i}{2}\left[\gamma^E_\mu,\gamma^E_\nu\right]\,.
\end{equation}

\subsection{Rare hyperon form factor decomposition in Euclidean spacetime} 
Schematically, the rare-hyperon decay matrix element in Minkowski-metric is given by
\begin{equation}
 \mathcal{A}^M_\mu=\overline{u}_p
 \widetilde{\mathcal{A}}^M_\mu u_{\Sigma} = \left<p\right| \mathcal H_W^M J_\mu^M\left|\Sigma\right>\,.
 \end{equation}
with the weak Hamiltonian given by $ \mathcal H_W^M \propto C_1 Q^M_1 + C_2 Q^M_2$. The operators $Q^M_1$ and $Q^M_2$ can be written as
\begin{align}
 Q_1^M&=\left[\overline{s}\gamma_\mu^{L,M}d\right]\left[\overline{q}\gamma_M^{L\,\mu}q\right]=\left[\overline{s}\gamma_\mu^{M}\left(1-\gamma_5^M\right)d\right]\left[\overline{q}\gamma_M^{\mu}\left(1-\gamma_5^M\right)q\right]\notag\\
 & = \left[\overline{s}\gamma_\mu^{E}\left(1+\gamma_5^E\right)d\right]\left[\overline{q}\gamma_{\mu}^E\left(1+\gamma_5^E\right)q\right]\equiv\left[\overline{s}\gamma_\mu^{L,E}d\right]\left[\overline{q}\gamma_\mu^{L,E}q\right]\equiv Q_1^E \,,
\end{align}
and
\begin{equation}
 Q_2^M=\left[\overline{s}\gamma_\mu^{L,M}q\right]\left[\overline{q}\gamma_M^{L\,\mu}d\right]=\left[\overline{s}\gamma_\mu^{L,E}q\right]\left[\overline{q}\gamma_\mu^{L,E}d\right]\equiv Q_2^E\,.
\end{equation}
For the electromagnetic current one finds
\begin{equation}
 J_\mu^M = \overline{q}\gamma_\mu^M q=\begin{cases}
   \overline{q}\gamma_0^M q  = \overline{q}\gamma_0^E q \equiv J_0^E \hspace{1.8cm}\text{for}\quad\mu=0\\[0.2cm] 
   \overline{q}\gamma_j^M q =-i\overline{q}\gamma_j^E q \equiv -iJ_j^E\hfill\text{for}\quad\mu=j
   \end{cases}
   \label{eq:emcurrE}
\end{equation}
where we define $J_\mu^E\equiv \overline{q}\gamma_\mu^E q$\,.
\\[0.4cm]
In total, for the rare-hyperon amplitude we obtain
\begin{equation}
 \mathcal{A}^M_\mu=\overline{u}_p
 \widetilde{\mathcal{A}}^M_\mu u_{\Sigma}=\left<p\right| \mathcal H_W^M J_\mu^M\left|\Sigma\right>=\begin{cases}
  \left<p\right| \mathcal H_W^E J_0^E\left|\Sigma\right> = \overline{u}_p
 \widetilde{\mathcal{A}}^E_0 u_{\Sigma}  \hspace{2.4cm}\text{for}\quad\mu=0\\[0.2cm]                                                                  
   \left<p\right| \mathcal H_W^E (-i)J_j^E\left|\Sigma\right> = \overline{u}_p
 (-i)\widetilde{\mathcal{A}}^E_j u_{\Sigma} \hfill\text{for}\quad\mu=j                                             \end{cases}
\end{equation}
i.e.\ we define
\begin{equation}
 {\mathcal{A}}^M_\mu \equiv \begin{cases}
   {\mathcal{A}}^E_0  \hspace{1.4cm}\text{for}\quad\mu=0\\[0.2cm]                                                                   
  -i {\mathcal{A}}^E_j \hfill\text{for}\quad\mu=j                                                                   \end{cases}
\end{equation}
As derived above, written with Minkowski quantities the form factor decomposition for the rare hyperon decay is given as
\begin{equation}
 \widetilde{\mathcal{A}}^M_\mu = i\sigma^M_{\nu\mu}q_M^\nu\left[a + \gamma^M_5 b\right] + \left(q^2_M\gamma_\mu^M-q_\mu^M\slashed{q}_M\right) \left[c + \gamma^M_5 d\right] \,.
\end{equation}
For $\mu=0$ one finds that this can be written with Euclidean quantities as follows:
\begin{align}
 \widetilde{\mathcal{A}}^M_0 & = i\sigma^M_{\nu0}q_M^\nu\left[a + \gamma^M_5 b\right] + \left(q^2_M\gamma_0^M-q_0^M\slashed{q}_M\right) \left[c + \gamma^M_5 d\right]\notag\\
 & = \sigma^E_{\nu0}q_E^\nu\left[a - \gamma^E_5 b\right] + \left(-q^2_E\gamma_0^E+q_0^E\slashed{q}_E\right) \left[c - \gamma^E_5 d\right]\equiv \widetilde{\mathcal{A}}^E_0\,.
 \label{eq:rareHA0}
\end{align}
Similarly, for $\mu=j$ can be written as 
\begin{align}
  \widetilde{\mathcal{A}}^M_j &= i\sigma^M_{\nu j}q_M^\nu\left[a + \gamma^M_5 b\right] + \left(q^2_M\gamma_j^M-q_j^M\slashed{q}_M\right) \left[c + \gamma^M_5 d\right]\notag\\
  & = i((-i\sigma^E_{0 j})(-iq_E^0)+(-\sigma^E_{ij})q_E^i)\left[a - \gamma^E_5 b\right] + \left(-q^2_E(-i\gamma_j^E)-(-q_j^E)(-i\slashed{q}_E)\right) \left[c - \gamma^E_5 d\right]\notag\\
    & = -i\sigma^E_{\nu j}q_E^\nu\left[a - \gamma^E_5 b\right] + i\left(q^2_E\gamma_j^E-q_j^E\slashed{q}_E\right) \left[c - \gamma^E_5 d\right]\equiv -i\widetilde{\mathcal{A}}^E_j\,.
\label{eq:rareHAj}
  \end{align}
In equations \eqref{eq:rareHA0} and \eqref{eq:rareHAj} we have defined
\begin{equation}
 \widetilde{\mathcal{A}}^E_\mu = \sigma^E_{\nu\mu}q_E^\nu\left[a - \gamma^E_5 b\right] + \left(-q^2_E\gamma_\mu^E+q_\mu^E\slashed{q}_E\right) \left[c - \gamma^E_5 d\right]\,,
\end{equation}
such that
\begin{equation}
 \mathcal{A}_\mu^E = \left<p\right| \mathcal H_W^EJ_\mu^E\left|\Sigma\right> = \overline{u}_p\widetilde{\mathcal{A}}^E_\mu u_{\Sigma}
\end{equation}
with the rare-hyperon decay matrix element written with Euclidean quantities.


%
%
%
%
%
\section{(Euclidean) traces for extracting the form factors}
\label{sec:traces}

All quantities ($\Gamma$-matrices etc) in this section are understood to be the Euclidean versions, and we drop sub-/superscripts ``E'' for better readability. Details of our conventions for Euclidean metric can be found in Appendix \ref{sec:appMinEuc} above.

In the following we give examples of traces and combinations that can be used to extract the form factors $a(q^2)$, $b(q^2)$, $c(q^2)$ and $d(q^2)$. For convenience we define
\begin{equation}
 \text{tr}^\Gamma_\mu \equiv  \text{Tr}\left[\Gamma (-i\slashed{p}_p+M_p) \tilde{\mathcal{A}}_\mu (-i\slashed{k}_{\Sigma^+}+M_{\Sigma^+})\right]
\end{equation}
with
\begin{equation}
 \tilde{\mathcal{A}}_\mu = \sigma_{\nu\mu}q^\nu\left[a - \gamma_5 b\right] + \left(-q^2\gamma_\mu+q_\mu\slashed{q}\right) \left[c - \gamma_5 d\right]\,.
\end{equation}
For example, one finds
\begin{equation}
\begin{aligned}
 \text{tr}^\mathds{1}_0 & = \text{Tr}\left[ (-i\slashed{p}_p+M_p) \tilde{\mathcal{A}}_0 (-i\slashed{k}_{\Sigma^+}+M_{\Sigma^+})\right] \\
 & = 4\left(a+c\left(M_p+M_{\Sigma^+}\right)\right)\cdot\left[E_{\Sigma^+} (\boldsymbol{p}\cdot\boldsymbol{q})-E_p (\boldsymbol{k}\cdot\boldsymbol{q})\right]
\end{aligned}
\end{equation}
or
\begin{equation}
\begin{aligned}
 \text{tr}^{\gamma_5}_0 & = \text{Tr}\left[\gamma_5 (-i\slashed{p}_p+M_p) \tilde{\mathcal{A}}_0 (-i\slashed{k}_{\Sigma^+}+M_{\Sigma^+})\right] \\
 & = 4\left(b+d\left(M_p-M_{\Sigma^+}\right)\right)\cdot\left[E_{\Sigma^+} (\boldsymbol{p}\cdot\boldsymbol{q})-E_p (\boldsymbol{k}\cdot\boldsymbol{q})\right]\,.
\end{aligned}
\end{equation}

The goal is to find suitable combinations of the traces $\text{tr}_\mu^\Gamma$ to extract the four form factors $a$, $b$, $c$ and $d$. As an example we will give results for the kinematics where the ${\Sigma^+}$ is at rest $\boldsymbol{k}=0$ and the proton is moving along the $x$-direction $\boldsymbol{p}=(p_p,0,0)$.\par

\begin{table}[ht]
  \begin{center}
 \begin{tabular}{|c|c|}
 \hline
  $\Gamma$ & $\text{tr}^\Gamma_0$\\
  \hline
  $\gamma_0$ & $-4 M_{\Sigma^+} (a+c (M_p+M_{\Sigma^+})) p_p^2$\\
  $\gamma_1$ & $4i M_{\Sigma^+} (a+c (M_p+M_{\Sigma^+})) p_p (E_p-M_p)$\\
    $\mathds{1}$ & $-4 M_{\Sigma^+} (a+c (M_p+M_{\Sigma^+})) p_p^2$\\
    $\gamma_5$ & $-4 (b+d (M_p-M_{\Sigma^+})) M_{\Sigma^+} p_p^2$\\
  $\gamma_0\gamma_5$ & $-4 (b+d (M_p-M_{\Sigma^+})) M_{\Sigma^+} p_p^2$\\  
  $\gamma_1\gamma_5$ & $4 i M_{\Sigma^+} (b+d (M_p-M_{\Sigma^+})) p_p (M_p+E_p)$\\
  $\sigma_{01}$ & $-4 M_{\Sigma^+} (a+c (M_p+M_{\Sigma^+})) p_p (E_p-M_p)$\\
  $\sigma_{23}$ & $-4 M_{\Sigma^+} (b+d (M_p-M_{\Sigma^+})) p_p (M_p+E_p)$\\
  \hline
 \end{tabular}
\caption{Traces for $\mu=0$ and $\vec{k}=0$, $\vec{p}=(p_p,0,0)$. All traces that are not explicitly given are $0$.}
\label{tab:traces0}
\end{center}
\end{table}
\begin{table}[ht]
  \begin{center}
 \begin{tabular}{|c|c|}
 \hline
  $\Gamma$ & $\text{tr}^\Gamma_2$\\
  \hline
$\gamma_2$ & $-4M_{\Sigma^+}(E_p-M_p)\left[a(M_p+M_{\Sigma^+})+c((E_p-M_{\Sigma^+})^2-p_p^2)\right]$\\
$\gamma_3$ & $-4iM_{\Sigma^+}\,p_p\left[ b(M_p-M_{\Sigma^+})+d ((E_p-M_{\Sigma^+})^2-p_p^2)\right]$\\
$\gamma_{2}\gamma_5$ & $4M_{\Sigma^+}(E_p+M_p)\left[ b(M_p-M_{\Sigma^+})+d ((E_p-M_{\Sigma^+})^2-p_p^2)\right]$\\
$\gamma_3\gamma_5$ & $4iM_{\Sigma^+}\,p_p\left[a(M_p+M_{\Sigma^+})+c((E_p-M_{\Sigma^+})^2-p_p^2)\right]$\\
 $\sigma_{02}$ & $4iM_{\Sigma^+}(E_p-M_p)\left[a(M_p+M_{\Sigma^+})+c((E_p-M_{\Sigma^+})^2-p_p^2)\right]$\\ 
 $\sigma_{03}$ & $4M_{\Sigma^+}\,p_p\left[ b(M_p-M_{\Sigma^+})+d ((E_p-M_{\Sigma^+})^2-p_p^2)\right]$\\ 
$\sigma_{12}$ & $-4M_{\Sigma^+}\,p_p\left[a(M_p+M_{\Sigma^+})+c((E_p-M_{\Sigma^+})^2-p_p^2)\right]$\\
$\sigma_{13}$ & $4iM_{\Sigma^+}(E_p+M_p)\left[ b(M_p-M_{\Sigma^+})+d ((E_p-M_{\Sigma^+})^2-p_p^2)\right]$\\
\hline
 \end{tabular}
\caption{Traces for $\mu=2$ and $\vec{k}=0$, $\vec{p}=(p_p,0,0)$. All traces that are not explicitly given are $0$.}
\label{tab:traces2}
\end{center}
\end{table}
\par

The (non-zero) traces for the amplitude using the temporal vector current $\mu=0$ are listed in table \ref{tab:traces0}. From these traces, one get extract  information on the following combinations of form factors:
\begin{equation}
 \text{tr}^\Gamma_0 \quad\Rightarrow\quad a+c(M_p+M_{\Sigma^+})\qquad\text{and}\qquad
b+d(M_p-M_{\Sigma^+})\,.
\end{equation}
 The same information is contained in $\text{tr}^\Gamma_1$ (when the proton momentum is in $x$-direction as in this example). Complementary information can be obtained from $\mu=2$ (and the same information from $\mu=3$) and the respective traces are listed in Table \ref{tab:traces2}. From these traces, one can extract values for the following combination of form factors:
 \begin{align}
 \text{tr}^\Gamma_2 \quad\Rightarrow\quad a(M_p+M_{\Sigma^+})+c((E_p-M_{\Sigma^+})^2-p_p^2)\qquad\text{and}\qquad
 b(M_p-M_{\Sigma^+})+d ((E_p-M_{\Sigma^+})^2-p_p^2)\,.
\end{align}
As an example\footnote{Using other combinations is possible as well; here we give only one of many possibilities.}, the following combinations of traces (and kinematic factors) can be used to extract the four form factors 
\begin{align}
 &a = \frac{M_{\Sigma^+}+M_p}{8M_{\Sigma^+}^2(M_p+E_p)p_p}\left[\text{tr}_0^{\mathds{1}}\frac{(E_p-M_{\Sigma^+})^2-p_p^2}{(M_{\Sigma^+}+M_p)p_p}-\frac{p_p}{(E_p-M_p)}\text{tr}_2^{\gamma_2}\right]\\[0.3cm]
 &c = \frac{1}{8M_{\Sigma^+}^2(M_p+E_p)p_p} \left[\frac{p_p}{(E_p-M_p)}\text{tr}_2^{\gamma_2} - \text{tr}_0^{\mathds{1}}\frac{(M_p+M_{\Sigma^+})}{p_p}\right] 
\end{align}
\begin{align}
 &b = \frac{M_{\Sigma^+}-M_p}{8M_{\Sigma^+}^2(E_p-M_p)p_p}\left[\text{tr}_0^{\gamma_5}\frac{(E_p-M_{\Sigma^+})^2-p_p^2}{(M_{\Sigma^+}-M_p)p_p} - \text{tr}_2^{\gamma_2\gamma_5}\frac{p_p}{E_p+M_p}\right]\\[0.3cm]
 &d = \frac{1}{8M_{\Sigma^+}^2(E_p-M_p)p_p}\left[\text{tr}_0^{\gamma_5}\frac{M_{\Sigma^+}-M_p}{p_p}-\text{tr}_2^{\gamma_2\gamma_5}\frac{p_p}{E_p+M_p}\right]\,.
\end{align}


%
%
%
%
%
%
%
\section{Details on extracting finite-volume excited states}
\label{app:details_fvstates}

We assume throughout that the $\Sigma^+$ is at rest in the finite-volume frame. Then the infinite-volume quantum numbers, $J^P=1/2^+$, of the $|\Sigma^+(\textbf{0})\rangle$ state, imply that it transforms in the $G_1^+$ irrep \cite{Johnson:1982yq,Basak:2005ir} of the double-cover of the octahedral group (with parity). In addition to the $J=1/2$ channel, this irrep couples to $J=7/2$, which corresponds to orbital angular momentum values of $\ell=3, 4$. Neglecting this mixing, by formally assuming that the $N \pi \to N \pi$ scattering amplitude vanishes for $\ell \geq 3$, the only sectors we have to consider are the $\ell=0$ ($P=-$) and $\ell=1$ ($P=+$) states with total angular momentum $J=1/2$. 

As discussed in the main text, the weak Hamiltonian density is a sum of a scalar ($\mathcal H_W^+$) and pseudo-scalar ($\mathcal H_W^-$) operators. The corresponding states $\mathcal{H}^+_W |\Sigma^+(\textbf{0})\rangle$ and $\mathcal{H}^-_W |\Sigma^+(\textbf{0})\rangle$ transform in the $G_1^+$ and $G_1^-$ irreps respectively and therefore couple to multi-hadron excited states with these quantum numbers. The latter correspond to different infinite-volume quantum numbers as follows:
\begin{align}
  G_1^-   \ : \ J=1/2, \ \ell=0, \ P=- \,, \\
 G_1^+ \  : \ J=1/2, \ \ell=1, \ P=+ \,.
\end{align}

With this information we can now construct the relevant states to remove from the amplitude. In both these irreps, the calculation is analogous 
to a phase-shift analysis for 
$J=1/2$ $N \pi$ scattering, similar to what has been done in \cite{Andersen:2017una,Andersen:2019ktw}. The operator basis is different in the two irreps and includes both $N \pi$ interpolators with different back-to-back momenta (for $G_1^-$, this includes an operator with both operators projected to zero spatial momentum, but for $G_1^+$, at least one unit of back-to-back momentum is required). Additionally, the Roper $N(1440)$ resonance is present in the $G_1^+$ irrep and a local quark bilinear corresponding to this might be included. In a more ambitious calculation three-hadron $N \pi \pi$ operators might also be incorporated.

Once the basis is established one can apply a suitable technique (e.g.~the method of distillation \cite{HadronSpectrum:2009krc,Morningstar:2011ka}) to form all possible two-point correlation functions from the operators, allowing one to construct a correlator matrix
\begin{align}
C^\Lambda(t,t_0) = 
\begin{pmatrix}
\langle O^\Lambda_1(t) O_1^{\Lambda\dagger} (t_0) \rangle &
\langle O^\Lambda_1 (t) O_2^{\Lambda\dagger} (t_0) \rangle &
\langle O^\Lambda_1 (t) O_3^{\Lambda\dagger} (t_0) \rangle & \cdots \\
\langle O^\Lambda_2 (t) O_1^{\Lambda\dagger}  (t_0) \rangle &
\langle O^\Lambda_2 (t) O_2^{\Lambda\dagger} (t_0) \rangle &
 \langle O^\Lambda_2 (t) O_3^{\Lambda\dagger} (t_0) \rangle & \cdots \\
\langle O^\Lambda_3 (t) O_1^{\Lambda\dagger}  (t_0) \rangle &
\langle O^\Lambda_3 (t) O_2^{\Lambda\dagger} (t_0) \rangle &
 \langle O^\Lambda_3 (t) O_3^{\Lambda\dagger} (t_0) \rangle & \cdots \\ 
\vdots & \vdots & \vdots & \ddots
\end{pmatrix} \,,
\end{align}
where $\Lambda \in \{G_1^-,G_1^+ \}$ denotes the irrep. 
In a practical implementation, this basis should include more operators than the number of finite-volume states one plans to reliably extract.
The next step is to apply the variational method, which requires one to solve a GEVP of the form \cite{Michael:1982gb,Michael:1985ne,Luscher:1990ck}
\begin{align}
C^\Lambda(t) v(t, t_0) = \lambda(t, t_0) C^\Lambda(t_0) v(t, t_0) \,.
\end{align}
From the eigenvalues one can then extract the energy spectrum of the system via
\begin{align}
\lambda_n(t, t_0) =  e^{-E^\Lambda_n(t-t_0)} + B e^{-\Delta E^\Lambda_n(t-t_0)} + \cdots
\, ,
\end{align} 
where, for a suitable choice of $t_0$, $\Delta E^\Lambda_n$ is an energy gap to the lowest level not covered by the operator basis \cite{Blossier:2009kd}. With the eigenvectors arising from the GEVP, an optimised interpolator
\begin{align}
X^\Lambda_n = v^\Lambda_n(t, t_0)^\dagger O^\Lambda \,,
\end{align}
can then be formed which couples particularly well to the $n^\mathrm{th}$ state in the system. 
These can then be used exactly as the proton creation operators in the main text, in order to extract the relevant energies and matrix elements both for determining $F_\mu(\boldsymbol p, \boldsymbol k)_L$ and the finite-volume corrections that define the amplitude.

%
%
%
%
%
%
%
\section{Wick contractions for the rare hyperon decay}
\label{sec:appWick}
In the following, we show the diagrams corresponding to the different Wick contractions of the four-point function $ \Gamma^{(4)}_\mu$ defined in eq.~\eqref{eq:fourptHJ}. There are six different contractions for each of the four topologies of the three-point function \eqref{eq:threeptH}.
\begin{figure}[h]
\centering
 \includegraphics[width=0.9\textwidth]{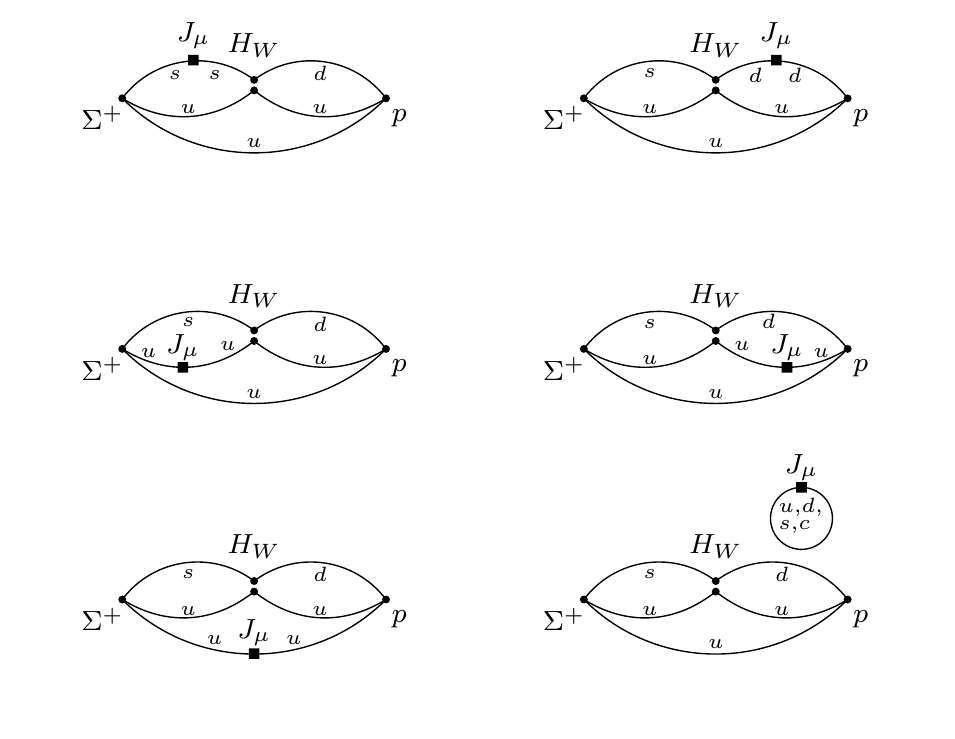}
 \caption{Four-point diagrams for the $C_{sd}$ class. The squared vertex shows the insertion of the electromagnetic current $J_\mu$.}
 \label{fig:Csd4pt}
\end{figure}
\newpage
\begin{figure}[h]
\centering
 \includegraphics[width=0.9\textwidth]{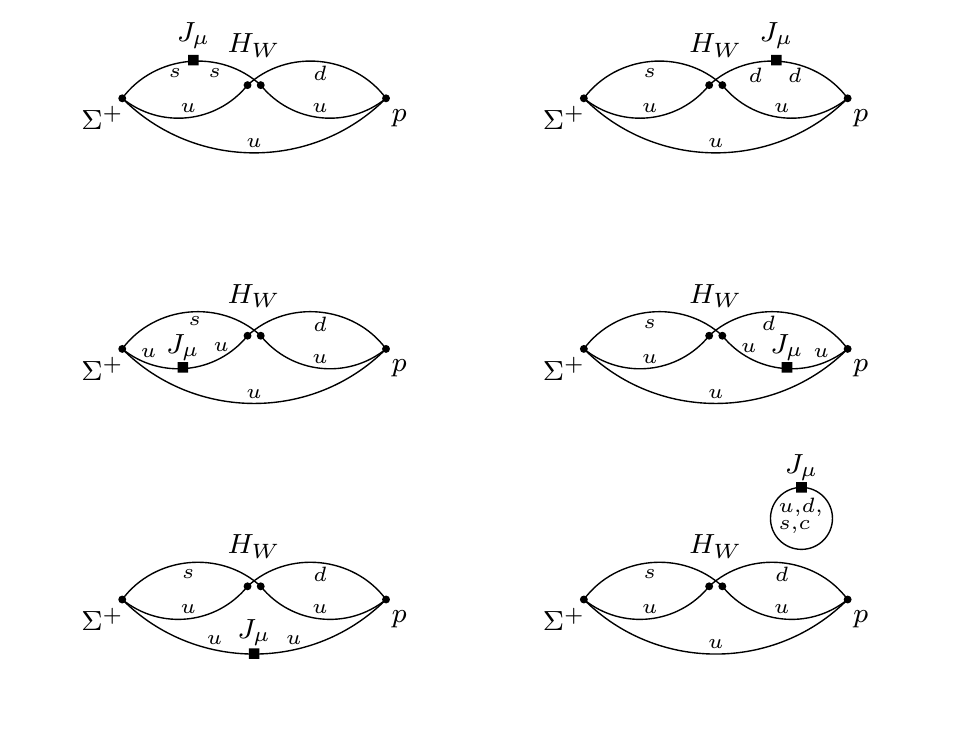}
 \caption{Four-point diagrams for the $C_{su}$ class. The squared vertex shows the insertion of the electromagnetic current $J_\mu$.}
  \label{fig:Csu4pt}
\end{figure}
\newpage
\begin{figure}[h]
\centering
 \includegraphics[width=0.9\textwidth]{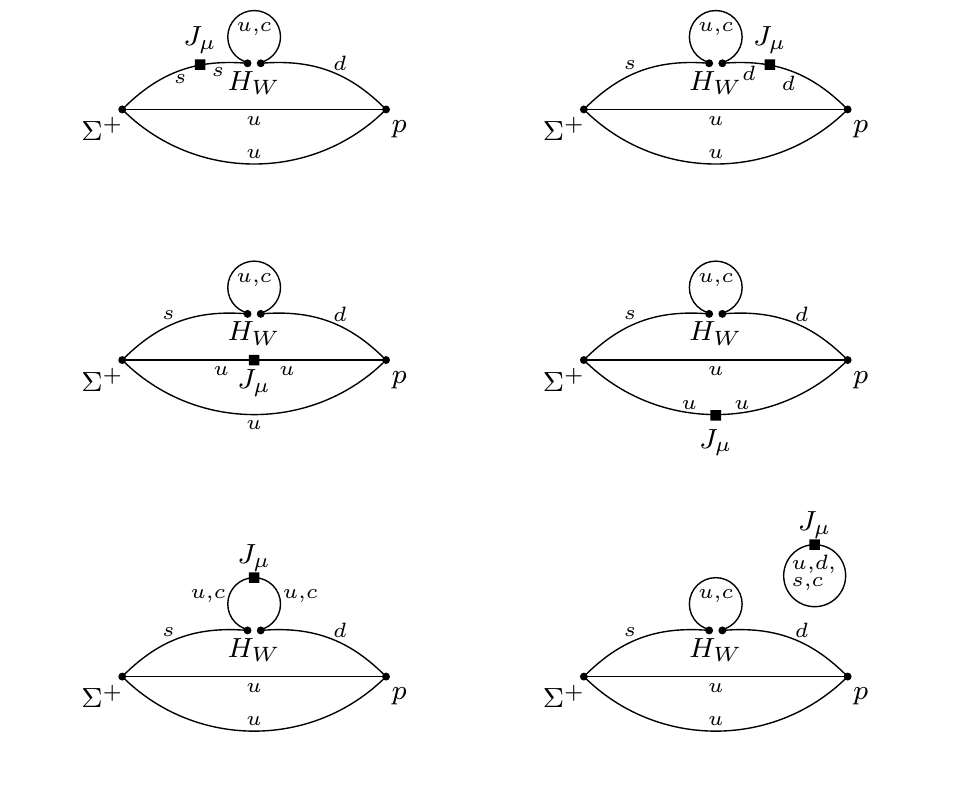}
 \caption{Four-point diagrams for the $S$ class. The squared vertex shows the insertion of the electromagnetic current $J_\mu$.}
  \label{fig:S4pt}
\end{figure}
\newpage
\begin{figure}[h]
\centering
 \includegraphics[width=0.9\textwidth]{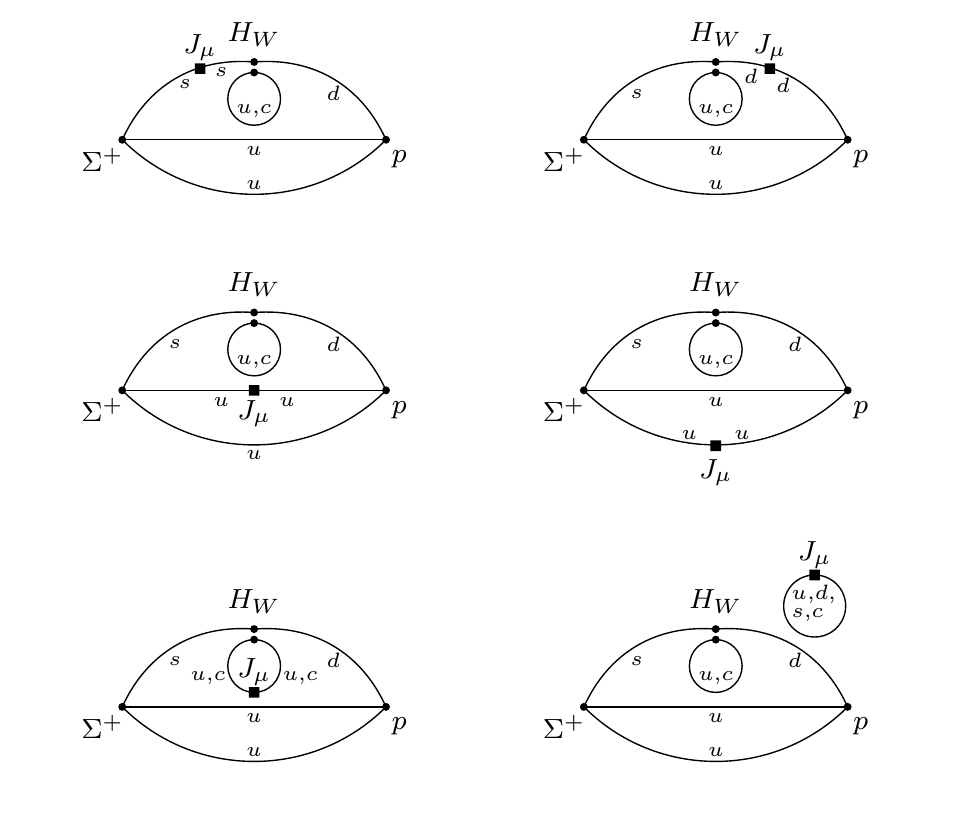}
 \caption{Four-point diagrams for the $E$ class. The squared vertex shows the insertion of the electromagnetic current $J_\mu$.}
  \label{fig:E4pt}
\end{figure}
%
%

\acknowledgments
A.P. would like to thank J.M. Camalich for initial discussions which prompted interest in this decay channel. F.E., V.G., R.H. and A.P. received funding from the European Research Council (ERC) under the European Union's Horizon 2020 research and innovation program under grant agreement No 757646 and A.P. additionally under grant agreement No 813942. M.T.H. is supported by UKRI Future Leaders Fellowship MR/T019956/1. Additionally, F.E., V.G., M.T.H. and A.P. are supported by UK STFC grant ST/P000630/1.

\bibliographystyle{JHEP}
\bibliography{rareHyperonDecay.bib}

\end{document}